\documentclass[pra,aps,twocolumn,amsmath,amssymb,floatfix,reprint,footinbib,superscriptaddress]{revtex4-2}
\usepackage{graphics}      % standard graphics specifications
\usepackage{graphicx}      % alternative graphics specifications
\usepackage{longtable}     % helps with long table options
\usepackage{url}           % for on-line citations
\usepackage{bm}    
\usepackage{braket}        % special 'bold-math' package
\usepackage{xcolor}
\usepackage{comment}
\usepackage{siunitx}
\usepackage[T1]{fontenc}
\usepackage[title]{appendix}
\usepackage{amsmath,amssymb,epsfig,color}
\usepackage{mathrsfs}
\usepackage{amsfonts}
\usepackage{dcolumn}
\usepackage{subfigure}
\usepackage{soul}
\usepackage{tabularx}
\usepackage[colorlinks,linkcolor=blue,citecolor=blue,hyperindex,bookmarks=false,pdfstartview=FitH]{hyperref}

\usepackage[ruled]{algorithm2e}  
\newcommand{\beq}{\begin{equation}}
\newcommand{\eeq}{\end{equation}}
\newcommand{\beqa}{\begin{eqnarray}}
\newcommand{\eeqa}{\end{eqnarray}}
\newcommand{\tr}{\text{Tr}}

\newcommand{\figref}[1]{\mbox{Fig.~\ref{#1}}}

\newcommand{\figpanel}[2]{\hyperref[#1]{\ref*{#1}(#2)}}
\newcommand{\figpanels}[3]{\hyperref[#1]{\ref*{#1}(#2)-(#3)}}

\newcommand{\figpanelNoPrefix}[2]{\hyperref[#1]{\ref*{#1}(#2)}}
\usepackage{mleftright} %to get rid of the extra space around parentheses with \left and \right: write \mleft, \mright instead 
 
\begin{document}

\title{Single-Period Floquet Control of Bosonic Codes with Quantum Lattice Gates}

\author{Tangyou Huang}
\author{Lei Du}
\email{lei.du@chalmers.se}
\affiliation{Department of Microtechnology and Nanoscience (MC2), Chalmers University of Technology, SE-41296 G\"oteborg, Sweden}

\author{Lingzhen Guo}
\email{lingzhen_guo@tju.edu.cn}
\affiliation{Center for Joint Quantum Studies and Department of Physics, School of Science, Tianjin University, Tianjin 300072, China}

\date{\today }

\begin{abstract}
Bosonic codes constitute a promising route to fault-tolerant quantum computing. {Existing Floquet protocols enable analytical construction of bosonic codes but typically rely on slow adiabatic ramps with thousands of driving periods.} In this work, we circumvent this bottleneck by introducing an analytical and deterministic Floquet method that directly synthesizes arbitrary unitaries within a single period. The phase-space unitary ensembles generated by our approach reproduce the Haar-random statistics, enabling practical pseudorandom states in continuous-variable systems.
We prepare various prototypical bosonic codes from vacuum and implement single-qubit logical gates with high fidelities using quantum lattice gates. By harnessing the full intrinsic nonlinearity of Josephson junctions, quantum lattice gates decompose quantum circuits into primitive operations for efficient continuous-variable quantum computing.
\end{abstract}

\maketitle

\textit{Introduction.---} Fault-tolerant quantum computation promises advantages over its classical counterpart, ranging from fundamental research to commercial applications. Practical fault tolerance typically relies on quantum error correction (QEC)~\cite{QEC96prl}, which has been experimentally demonstrated across discrete-variables (DV) and continuous-variables (CV) platforms~\cite{google2023suppressing,sivak2023real}, as well as in hybrid DV--CV architectures~\cite{QEC2024cat}. 
Unlike the DV cases, CV protocols leverage the infinite-dimensional Hilbert space of harmonic oscillators, offering redundancy in encoding quantum information without the need for large numbers of physical qubits~\cite{CV2005RMP,CV2012RMP}.

{A universal set of single-mode CV quantum gates can be realized using Gaussian operations, such as displacements, squeezing, and phase rotations, in combination with a single nonlinear operation such as the cubic phase gate~\cite{Lloyd1999prl}.} In superconducting circuits, the cubic phase gate has been demonstrated using an on-chip planar resonator terminated by a SNAIL element~\cite{Hillmann2020SNAIL,Eriksson2024SNAIL}. Another popular single-mode universal gate set combines displacements with the selective number-dependent arbitrary phase (SNAP) gates, which impart arbitrary phases to individual Fock states via an off-resonantly coupled ancilla transmon qubit~\cite{Heeres2015SNAP,Liang2015praR,heeres2017snap}. The universal gate set for multimode CV quantum computation requires only the inclusion of a simple two-mode linear coupling~\cite{Sefi2011twomode,Budinger2024twomode}. 
However, universality alone is not sufficient for fault-tolerant quantum computation. On the one hand, the continuous nature of bosonic-quadrature errors necessitates embedding a finite-dimensional code space within the infinite-dimensional CV Hilbert space~\cite{Milburn2002Encode}. {On the other hand,  non-Gaussian resources are typically required to correct Gaussian errors and implement fault-tolerant quantum computation~\cite{Larsen2025Nature,Konno2024Science}.} When the encoded states satisfy the Knill–Laflamme conditions~\cite{Knill97pra}, they realize bosonic QEC codes~\cite{BCreviewJoshi,BCreviewSun,BCreviewJiang}.

Both the cubic phase gate and the SNAP gate need complicated design of superconducting circuits using the low-order nonlinearity of Josephson junctions (JJs)~\cite{rojkov2024}.
In contrast, the recently introduced \textit{quantum lattice gates} (QLGs) leverage the full nonlinearity of JJ superconducting circuits and can provide an analytical framework based on the Floquet engineering to prepare and manipulate bosonic codes ~\cite{LZGuo2025CP}.
However, this approach suffers from slow adiabatic ramps~\cite{gkp2024prl,LZGuo2025CP} costing thousands of Floquet periods, which severely limits the speed and practical scalability for universal quantum computation.

{
In this work, we overcome this bottleneck by introducing an analytical, deterministic Floquet-engineering method that directly synthesizes arbitrary bosonic unitaries {from an arbitrary initial state} within a \textit{single} Floquet period. We validate our method by generating gate ensembles whose fidelity statistics reproduce those of Haar-random states~\cite{Lloyd2003pseudo-random}. {The numerical results show that, for Haar-state preparation, the computational resources required by our method scale linearly with the Hilbert-space dimension, with a substantially smaller prefactor than the SNAP-based approach.}
We further develop an \textit{optimal pulse engineering} (OPE) technique to prepare binomial~\cite{Binomial2016prx}, cat~\cite{Cat1999pra,QEC2024cat}, and Gottesman–Kitaev–Preskill (GKP)~\cite{gkp2001pra} codes directly from vacuum, achieving state infidelities below $10^{-5}$.
Furthermore, we demonstrate high-fidelity logical gates for these representative bosonic codes, with average gate errors on the order of $10^{-4}$, in \emph{few-microsecond} QLGs, which is over three orders of magnitude faster than existing adiabatic protocols~\cite{gkp2024prl,guo2024prl,LZGuo2025CP}.}

\textit{Noncommutative Floquet Engineering.---} From the Floquet theory, the stroboscopic evolution over one period of a time-periodic Hamiltonian $\hat{H}(t)=\hat{H}(t+T)$ is captured by a time-independent Floquet Hamiltonian $\hat{H}_{\rm F}$~\cite{Shirley1965pr,Eckardt2015NJP,Liang2018njp},
\beqa\label{eq:floquet}
\exp\Big(-i\frac{1}{\lambda}\hat{H}_{\rm F}T\Big)\equiv\mathcal{T}\exp\Big[-\frac{i}{\lambda}\int_{0}^{T}\hat{H}(t)dt\Big], 
\eeqa
where {$\lambda$ is an effective dimensionless Planck
constant, $T$ is the Floquet period, and} $\mathcal{T}$ is the time-ordering operator.
In this work, we consider a driven cavity described by $\hat{H}_{\rm cav} =\hbar\omega_0 \hat{a}^\dagger \hat{a} + V(t)$ \cite{guo2024prl}, where $\hat{a}$ ($\hat{a}^{\dag}$) is the annihilation (creation) operator of the cavity.
Our objective is to realize an arbitrary unitary operation on a $d$-dimensional subspace of cavity mode.
For any target unitary operator $\hat{U}_{\rm tar}\in \mathbb{U}(d)$, there exists a Hermitian generator \( \hat{H}_{\rm tar} \) such that {$\hat{U}_{\rm tar} = e^{-i \hat{H}_{\rm tar}/\lambda}$}.
The key idea is to design a periodic potential $V(t)$ for synthesizing an arbitrary Floquet Hamiltonian in the cavity-frequency rotating frame $\hat{H}_{\rm tar} = \sum_{j,l}  \epsilon_{j,l}(\hat{a}^\dagger)^{j} \hat{a}^{l}$, where $\epsilon_{j,l}=\epsilon_{l,j}^*\in\mathbb{C}$ denote the complex driving amplitudes. 
We write the target Hamiltonian in the orthonormal Fock basis \(\{|n\rangle\}_{n=0}^{d-1}\) as
$\hat{H}_{\rm tar} = \sum_{n,m=0}^{d-1} h_{n,m} |n\rangle \langle m| $
with $h_{n,m}= \langle n|\hat{H}_{\rm tar} |m\rangle$. 

According to the technique of \emph{noncommutative Fourier transformation} (NcFT)~\cite{guo2024prl}, the target Hamiltonian is decomposed as a superposition of plane-wave operators, 
\beq\label{eq-HTxp}
\hat{H}_{\rm tar}(\hat{x},\hat{p})
=
\frac{1}{2\pi}\int \int dk_x dk_pf_{\rm tar}(k_x,k_p)e^{i(k_x\hat{x}+k_p\hat{p})}.
\eeq
Here, the quadrature operators are defined as $\hat{x}\equiv\sqrt{\frac{\lambda}{2}}(\hat{a}^{\dag}+\hat{a})$ and $\hat{p}\equiv i\sqrt{\frac{\lambda}{2}}(\hat{a}^{\dag}-\hat{a})$, which satisfy $[\hat{x},\hat{p}]=i\lambda$. {$k_x$ and $k_p$ are phase-space Fourier variables associated with quadratures $\hat{x}$ and $\hat{p}$, and $e^{i(k_x\hat{x}+k_p\hat{p})}$ can be viewed as a phase-space displacement generator.} By introducing the variables $k=\sqrt{k_{x}^{2}+k_{p}^{2}}$ and $\tau\equiv\omega_0 t=\mathrm{Arg}(k_{x}+ik_{p})$, the NcFT coefficient $f_{\rm tar}$ is given in the Fock basis by
\beq
f_{\rm tar}(k,\tau)=\sum_{n,m}h_{n,m}f_{n,m}(k,\tau).
\label{ftar}
\eeq
The explicit kernels $f_{n,m}(k,\tau)$ are given in the Supplemental Material (SM)~\cite{SM}.
With the NcFT coefficient, the synthesized driving potential can be written as a superposition of cosine-type potentials~\cite{guo2024prl}
\beqa\label{eq:exp-cos}
\hat{V}(\hat{x},\tau)
&=&\int_{-\infty}^{+\infty}\beta_0A(k, \tau)\cos[k\hat{x}+\phi(k,\tau)]\,dk,
\eeqa 
where the target-independent parameter $\beta_0$ controls the driving strength. The time-dependent amplitude $A(k,\tau)$ and phase $\phi(k,\tau)$ are given by
\beqa\label{eq-Aphi}
\left \{ \begin{array}{lll}
	A(k,\tau)&=&k\Big|f_{\rm tar} (k\cos{\tau},k\sin{\tau})\Big|,\\
	\phi(k,\tau)&=&\text{Arg}\Big[f_{\rm tar} (k\cos{\tau},k\sin{\tau})\Big].
\end{array} \right.
\eeqa
{As shown in Eq.~\eqref{eq:exp-cos}, the target potential is synthesized by superposing the cosine-type components with $A(k,\tau)$ and $\phi(k,\tau)$. Experimentally,} each cosine component can be physically implemented, for example, with optical lattices in ultracold atom setups~\cite{Moritz2003prl}, or JJ potentials in superconducting circuits~\cite{Hofheinz2011prl}.  
%
%{\color{blue} See \cite{SM} for more details on experimental implementations. In this way, we reduce the synthesis of an arbitrary unitary to the design of a periodic driving Hamiltonian via Floquet engineering. }
%
To the leading order in the Floquet–Magnus (FM) expansion, the target unitary operator is given by
\begin{equation}
\hat{U}_{\rm tar} \approx   \mathcal{T}\exp\!\left[-\frac{i}{\lambda}\int_{0}^{T}\hat{V}(\tau)\,d\tau\right].
\label{eq:Uapprox}
\end{equation}
In principle, higher-order FM errors can be mitigated by adding driving terms perturbatively~\cite{guo2024perturbative}.

\begin{figure} 
\centering
\includegraphics[width=\columnwidth]{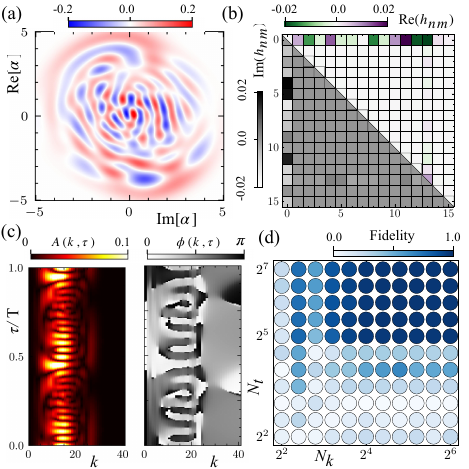}%width=\columnwidth scale=0.3
\caption{\textbf{{Single-period Floquet control} with quantum lattice gates (QLGs).} 
(a) Wigner function of the target state $|\psi_{\rm tar}\rangle$. (b) Target Hamiltonian $\hat{H}_{\rm tar}= i\lambda\log(\hat{U}_{\rm tar})$ in the Fock basis, where the target unitary $\hat{U}_{\rm tar}$ is constructed from Eq.~(\ref{eq:unitary learning}). (c) Amplitude $A(k,\tau)$ and phase $\phi(k,\tau)$ of the driving potential over one Floquet period $T$. 
(d) State-preparation fidelity $\mathcal{F}=|\bra{\psi_{\rm tar}} \hat{U}_{\rm tar} |\psi_0\rangle|^2$ as a function of Trotter depths $N_k$ and $N_t$. We use $N_t = 2^6$ and $N_k=40$ in (a-c). Other parameters: $N_p=15$, $k_f=40$, and $\beta_0=1$. 
%\htyr{Note: We have corrected the y-axis label in (c), i.e., $t\to\tau$.}
}
\label{fig:figure1}
\end{figure}

\textit{Quantum lattice gate.---}
{In our protocol, we approximate the target unitary [Eq.\eqref{eq:Uapprox}] by a Trotter–Suzuki (TS) decomposition of the driving potential [Eq.\eqref{eq:exp-cos}], which leads to a sequence of QLGs~\cite{LZGuo2025CP},}
\beqa
\hat{\mathbf{G}}(\vec{\theta},\vec{\gamma}) \equiv \prod_{n =1}^{N_t }\prod_{ m =1}^{ N_k} \hat{\mathbf{g}}(\theta_{n,m}, \gamma_{n,m})
\label{eq:Trotter_error}
\eeqa
with $N_{t}$ ($N_{k}$) the number of discretization steps, i.e., the Trotter depth, for time (wavenumber), and $\hat{\mathbf{g}}(\theta_{n,m}, \gamma_{n,m})$ the elementary QLG defined via
\beqa\label{eq:QLG}
\hat{\mathbf{g}}(\theta_{n,m}, \gamma_{n,m})\equiv e^{-\frac{i}{\lambda}\theta_{n,m}\cos( k_n \hat{x} + \gamma_{nm})}.
\eeqa
Here, we have introduced parameters  $\gamma_{n,m} = \phi(k_n,\tau_m)$, $\theta_{n,m} = \beta_0A(k_n,\tau_m) (T/N_t)(k_f/N_k)$ with $\tau_{m}=m(T/N_t)$, $k_{n}=n(k_f/N_k)$, where $k_{f}$ is an appropriate cutoff for the wavenumber.
Except for errors from the high-order FM expansion,
another error of $\hat{\mathbf{G}}(\vec{\theta},\vec{\gamma})$ approximating the target unitary $\hat{U}_{\rm tar}$ originates from the TS decomposition [cf. Eq.~\eqref{eq:Trotter_error}], which can be suppressed by employing higher-order TS expansions and adaptive optimization strategies~\cite{Trotter1,trotterAdaptive,trotterAdaptive1}. {In this way, an arbitrary target unitary $U_{\rm tar}$ acting on a truncated $d$-dimensional Fock space can be synthesized, within a single Floquet period, by a sequence of QLGs.}

{\textit{Single-period Floquet control with QLGs.---}}
We first apply our protocol to prepare a target state $|\psi_{\rm tar}\rangle$ from a known initial state $|\psi_{0}\rangle$ via a unitary, i.e., $|\psi_{\rm tar}\rangle =\hat{U}_{\rm tar}\ket{\psi_0}$. We exclude the trivial case $\ket{\psi_{\rm tar}}= e^{i\eta}\ket{\psi_0}$ for some global phase $\eta$. {Previous approaches based on adiabatic ramping suffer from intrinsic limitations: maintaining adiabaticity requires a sufficiently slow Hamiltonian variation, which typically results in evolution over thousands of Floquet periods~\cite{guo2024prl,gkp2024prl,LZGuo2025CP}.}
In this work, we use a Householder-based construction~\cite{qHouseholder,Urias2010} to obtain a closed-form unitary that maps the known initial state $|\psi_{0}\rangle$ to the target state $|\psi_{\rm tar}\rangle$, without the need for any adiabatic processes. {In the supplemental material~\cite{SM}, we provide details on the Householder-based construction and discuss alternative unitary construction methods.} By defining the normalized Householder vector $\ket{u}\equiv(\ket{\psi_0}-e^{i\varphi}\ket{\psi_{\rm tar}})/\left\lVert \ket{\psi_0}-e^{i\varphi}\ket{\psi_{\rm tar}} \right\rVert$ with $e^{i\varphi}\equiv\braket{\psi_{\rm tar}|\psi_0}/|\braket{\psi_{\rm tar}|\psi_0}|$ (setting $e^{i\varphi}=1$ if $\braket{\psi_{\rm tar}|\psi_0}=0$), the target unitary can be obtained explicitly as
\beq\label{eq:unitary learning}
\hat{U}_{\rm tar}=\Big[\hat{I}+\big(e^{-i\varphi}-1\big)\ket{\psi_{\rm tar}}\!\bra{\psi_{\rm tar}}\Big]\Big(\hat{I}-2\ket{u}\!\bra{u}\Big),
\eeq
where $\hat{I}$ is the identity operator. It is straightforward to verify that $\hat{U}_{\rm tar}|\psi_{0}\rangle = |\psi_{\rm tar}\rangle$, while $\hat{U}_{\rm tar}$ acts as the identity (up to an overall phase) on the subspace orthogonal to the space spanned by $\ket{\psi_0}$ and $\ket{\psi_{\rm tar}}$.
Such $\hat{U}_{\rm tar} = e^{-i \hat{H}_{\rm tar}/\lambda}$ can then be decomposed into a sequence of elementary QLGs using the Floquet Hamiltonian engineering described above.

In \figref{fig:figure1}, we illustrate the workflow for preparing arbitrary target states $\ket{\psi_{\rm tar}}$ from the vacuum $\ket{\psi_0}=\ket{0}$ using QLGs. Starting from a randomly chosen nonclassical state, with its Wigner function shown in Fig.~\figpanel{fig:figure1}{a}, we first construct the target unitary $\hat{U}_{\rm tar}$ using the unitary-construction scheme, cf. Eq.~\eqref{eq:unitary learning}, 
%such that $\ket{\psi_{\rm tar}} \equiv  \hat{U}_{\rm tar}\ket{0}$ 
with arbitrary high fidelity $\mathcal{F} = |\bra{\psi_{\rm tar}} \hat{U}_{\rm tar} |\psi_0\rangle|^2$ (limited only by the machine precision in our simulation)~\cite{SM}. The corresponding target Hamiltonian obtained via $\hat{H}_{\rm tar} = i\lambda\log(\hat{U}_{\rm tar}) \equiv \sum_{n,m}h_{n,m}\ket{n}\bra{m}$ is visualized in Fig.~\figpanel{fig:figure1}{b}.
Next, from the NcFT representation of $\hat{H}_{\rm tar}$, cf. Eq.~(\ref{eq-HTxp}-\ref{eq:exp-cos}), we deterministically obtain the time-dependent driving amplitude and phase $A(k,\tau)$ and $\phi(k,\tau)$, as shown in Fig.~\figpanel{fig:figure1}{c}.  
Finally, the driving potential is decomposed into a sequence of QLG operations given by Eq.~(\ref{eq:Trotter_error}). Figure~\figpanel{fig:figure1}{d} shows how the state-preparation fidelity depends on the discretization parameters \(\{N_t, N_k\}\). For the example shown, our approach prepares the target states with fidelity $\gtrsim 0.99$ using $N_t=2^6$, $N_k=40$ and $d=15$. Moreover, Fig.~\figpanel{fig:figure1}{d} shows that the Trotter error associated with $N_k$ is less pronounced than the error from the time slicing $N_t$, we vary $N_t$ while holding $N_k$ fixed for subsequent benchmarks. Note that the entire preparation is completed within a single Floquet period, indicating that our approach is at least three orders of magnitude faster
than the adiabatic ramp method~\cite{gkp2024prl,guo2024prl,LZGuo2025CP}. {A detailed comparison of our method with previous protocols is provided in the End Matter.}

\begin{figure} 
\centering
\includegraphics[width=\columnwidth]{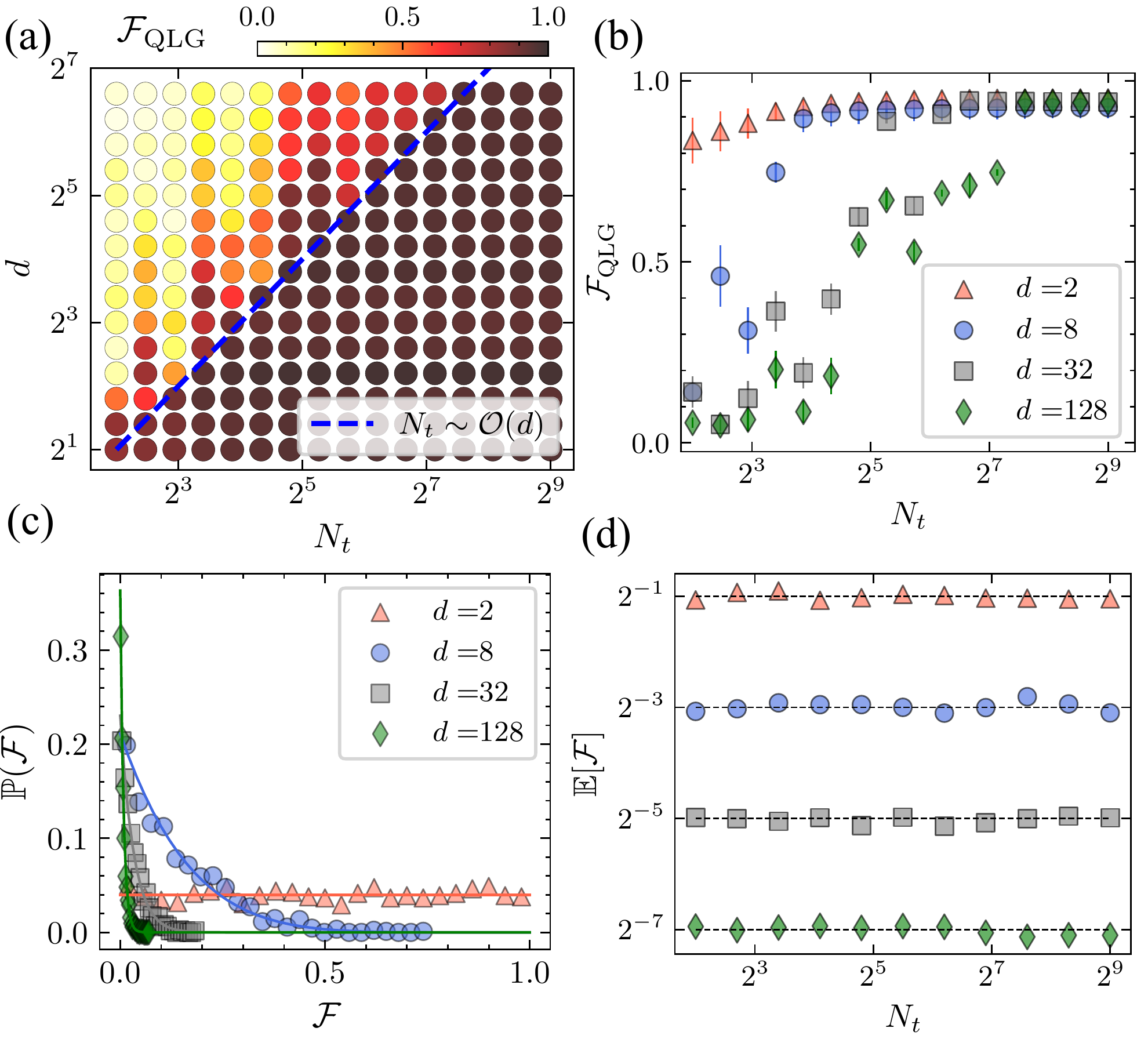}%width=\columnwidth scale=0.3
\caption{\textbf{Haar-random state benchmark.}  
(a) Average fidelity $\mathcal{F}_{\rm QLG}\equiv|\langle \psi_{\rm QLG} | \psi_{\rm Haar} \rangle|^2$ of Haar-state preparation  versus Trotter depth $N_t$ and Hilbert subspace dimension $d$, where each marker is averaged over $10^2$ randomly sampled Haar unitaries. {The blue dashed line indicates the empirical scaling $N_t \sim \mathcal{O}(d)$ required to reach high-fidelity ($\mathcal{F}_{\rm QLG}>0.95$)}. 
(b) Fidelity $\mathcal{F}_{\rm QLG}$ versus $N_t$ for different dimension $d$ ($10^3$ samples per point).
(c) Probability density distribution $\mathbb{P}(\mathcal{F})$ of the stochastic fidelity $\mathcal{F}=|\langle \psi_{\rm QLG} | \psi_{\rm ref} \rangle|^2$ over $10^2\times d$ from QLG realizations (markers) compared with the analytical Haar statistics $\mathbb{P}_{\rm Haar}(\mathcal{F})$ given by Eq.~(
\ref{eq:Haar}
) (solid curves) for different dimensions.   (d) Mean fidelity $\mathbb{E}[\mathcal{F}]$ versus $N_t$, with the theoretical prediction $1/d$ shown as dashed lines. Other parameters are the same as in \figref{fig:figure1}.}
\label{fig:figure2}
\end{figure}

{\textit{Haar-random state benchmark.---}
Haar-typical states and unitaries serve as a widely used benchmark for assessing quantum controllability in high-dimensional Hilbert spaces. 
In this context, we sample targets from the Haar measure and examine whether the implemented operations faithfully reproduce the statistics of Haar randomness~\cite{HaarPDF}. 
Such pseudorandom behavior underlies a broad range pf quantum information protocols and can be realized using gate-based quantum circuits~\cite{Hsin-Yuan2025random} and CV schemes involving non-Gaussian resources~\cite{PhysRevX.14.011013,dy4m-gq5c}.

Here, we construct a QLG-based unitary generator on a $d$-dimensional subspace whose output matches Haar statistics in phase space. Concretely, we define Haar-random states $|\psi_{\rm Haar}\rangle\equiv\hat{U}_{\rm Haar}|0\rangle$ drawn from the Haar measure on the unit sphere in \(\mathbb{C}^d\)~\cite{SM}, and study the distribution of fidelities with respect to a reference Haar state \( |\psi_{\rm ref}\rangle \). For an ensemble of Haar-distributed pure states, the fidelity $\mathcal{F}=|\langle \psi_{\rm Haar} | \psi_{\rm ref} \rangle|^2$ is a variable following the probability density function (PDF)~\cite{HaarPDF}
\beq\label{eq:Haar}
\mathbb{P}_{\rm Haar}(\mathcal{F}) = (d-1)(1 - \mathcal{F})^{d-2},
\eeq
with the mean fidelity $\mathbb{E}[\mathcal{F}] = 1/d$. In our implementation, we take each sampled $|\psi_{\rm Haar}\rangle$ as the target state and construct the target unitary $\hat{U}_{\rm Haar}$ from Eq.~(\ref{eq:unitary learning}) with the initial vacuum state $|0\rangle$.
We then design a sequence of QLGs from Eq.~(\ref{eq:Trotter_error}) for each target unitary $\hat{U}_{\rm Haar}$ yielding
\beq\label{eq:GHaar}
|\psi_{\rm QLG}\rangle = \hat{\mathbf{G}}_{\rm Haar}(\vec{\theta},\vec{\gamma})|0\rangle,
\eeq
which approximates the sampled Haar state  $|\psi_{\rm Haar}\rangle$ with high fidelity $\mathcal{F}_{\rm QLG}\equiv|\langle \psi_{\rm QLG} | \psi_{\rm Haar} \rangle|^2$.

Figure~\figpanel{fig:figure2}{a} shows the average fidelity $\mathcal{F}_{\rm QLG}$ of Haar-state preparation using QLGs as a function of the Hilbert subspace dimension $d$ and the Trotter depth $N_t$. {The blue dashed line indicates the empirical scaling $N_t \sim \mathcal{O}(d)$ required to reach high fidelity ($\mathcal{F}_{\rm QLG}\gtrsim 0.95$). As further illustrated by Fig.~\figpanel{fig:figure2}{b}, choosing $N_t = 2d$ yields $\mathcal{F}_{\rm QLG}\gtrsim 0.98$ for $d = 2, 8, 32, 128$.}
This overhead can be further reduced by higher-order TS formulas~\cite{PF2021prx} and numerical optimization~\cite{trotterAdaptive} which we leave for future work.  
In Fig.~\figpanel{fig:figure2}{c}, we 
calculate the fidelity $\mathcal{F}=|\langle \psi_{\rm QLG} | \psi_{\rm ref} \rangle|^2$ with respect to a fixed reference Haar state and plot the probability density distributions $\mathbb{P}(\mathcal{F})$ for different subspace dimensions, which are consistent with the Haar statistics given by Eq.~\eqref{eq:Haar}. {Figure~\figpanel{fig:figure2}{d} further shows the mean fidelity as a function of $N_t$. All curves approach the theoretical value $1/d$ with only weak dependence on the Trotter depth, indicating that the QLG protocol preserves the Haar statistics.}
These results show that our QLG approach reproduces Haar-like statistics faithfully, and thus can serve as a practical generator of pseudo-random states. {We also benchmark our method against a SNAP-gate-based construction in the End Matter.}

\begin{figure} 
\centering
\includegraphics[width=\columnwidth]{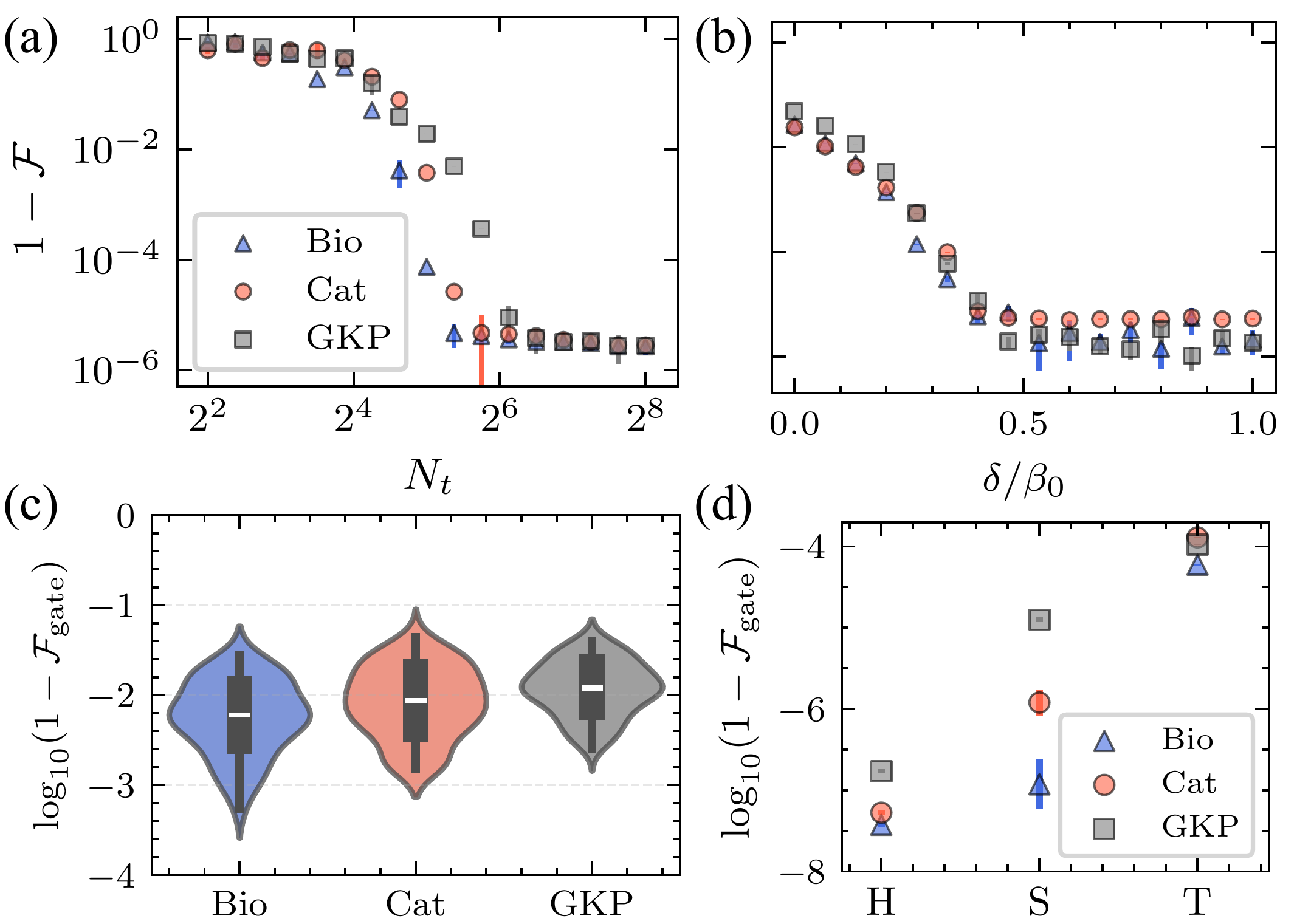}%width=\columnwidth scale=0.3
\caption{
\textbf{Bosonic code preparation and operation.}
(a)-(b): Infidelity of prepared code states versus (a) Trotter depth $N_t$ and (b) control bound $\delta / \beta_0$ for the binomial (triangle), cat (circle), and GKP (square) codes. (c) Violin plots for the binomial (Bio), Cat and GKP codes by randomly sampling over $10^2$ logical gates. The width of each violin (colored region) for a given infidelity logarithm value $\log_{10}(1 - \mathcal{F}_{\mathrm{gate}})$ is proportional to the probability density of the infidelity. The white line marks the median and the black box represents the interquartile range. (d) Infidelity of elementary single-qubit gates $\{ {\tt H}, {\tt S}, {\tt T} \}$ realized via quantum optimal control for each code. 
The error bars indicate standard deviation over $10^2$ data points.
Parameters: $\delta = 1$ for (a), $N_t = 2^6$ for (b), and $N_t = 2^8$, $\delta = 2$ for (c,d). Other parameters follow \figref{fig:figure1}.
}
\label{fig:figure3}
\end{figure}

\textit{Bosonic codes preparation and logical operations.---}   
We next validate our QLG approach by demonstrating the preparation of binomial~\cite{hu2019binomial}, cat~\cite{QEC2024cat}, and GKP codes~\cite{gkp2001pra}, see a brief introduction to these prototypical bosonic codewords in the SM~\cite{SM}. Here, we focus on the preparation of the logical zero code states, i.e., the binomial code $|\psi_{\rm bin}\rangle = (|0\rangle+\sqrt{3}|4\rangle)/2$ and the
four-component cat code   
$|\psi_{\rm cat}\rangle =\frac{1}{\sqrt{\mathcal{N}_{\rm cat}}}\big(|\alpha\rangle+|-\alpha\rangle+|i\alpha\rangle+|-i\alpha\rangle\big)$ with (real) coherent-state amplitude $\alpha$ and normalization factor $\mathcal{N}_{\rm cat}=8e^{-\alpha^2}[\cosh\alpha^2+\cos\alpha^2]$. Hereafter we choose the second sweet spot $\alpha=2.3447$ for the cat state, which satisfies Knill-Laflamme condition $\tan\alpha^2=\tanh \alpha^2$~\cite{Knill97pra,Knillooprl,SM}.
% And the  finite GKP state \cite{Simone2022prxq}: $|\psi_{\rm GKP}\rangle = \mathcal{N}_{\rm GKP} \sum_{s \in \mathbb{Z}} e^{-2\pi \Delta^2 s^2} \, e^{ -\frac{(q - 2s\sqrt{\pi})^2}{2\Delta^2} }$
For the finite-energy GKP code parameterized by a real width parameter $\sigma \in [0,1]$, we use its Fock-basis form  
%\beq\label{eq:finiteGKP}
$\ket{\psi_{\mathrm{GKP}}} = \sum_{\alpha' \in \mathcal{K}(\mu)} 
e^{-\sigma^2 |\alpha'|^2} e^{-i \mathrm{Re}[\alpha'] \mathrm{Im}[\alpha']} 
\hat{\mathcal{D}}(\alpha') \ket{0}$,
%\eeq
where $\hat{\mathcal{D}}(\alpha')$ is the displacement operator. The complex amplitude $\alpha'$ is chosen from the discrete grid  
$\mathcal{K}(\mu) = \left(\sqrt{\frac{\pi}{2}}\right) (2n_1 + \mu) + i \left(\sqrt{\frac{\pi}{2}}\right) n_2,$
with $\mu \in \{0,1\}$ the logical index, and $(n_1, n_2)$ spanning a finite integer range. The Gaussian envelope $e^{-\sigma^2|\alpha'|^2}$ imposes a finite-energy constraint, while the phase factor $e^{-i \mathrm{Re}[\alpha'] \mathrm{Im}[\alpha']}$ ensures the correct symplectic structure in phase space~\cite{SM}.  

Moreover, our approach enables single-qubit logical gate operations on code space spanned by the two orthogonal 
bosonic code states. In particular, arbitrary target single-qubit unitaries $\hat{\mathcal{G}}_{\rm tar}\in \mathbb{SU}(2)$ on the two-dimensional logical space can be synthesized via QLGs for the different bosonic codes introduced above. According to the \textit{Solovay--Kitaev theorem}~\cite{dawson2005solovay}, any single-qubit unitary can be $\varepsilon$-approximated using $O(\log^c(1/\varepsilon))$ gates with a constant $c\approx 2$.
For instance, the set of Hadamard ($\tt H$), phase ($\tt S$) and $\pi/8$~($\tt T$) gates constitutes a standard basis for universal quantum computation; within our framework these gates compile to QLGs for all three codes mentioned above.
% ; 
% \red{fault-tolerant implementations of these gates are available for several bosonic codes (code-dependent).}
%
To quantify the performance of the QLGs-compiled single-qubit gate $\hat{\mathcal{G}}_{\rm QLG}$, we benchmark on the logical subspace using the average gate fidelity defined by \cite{nielsen2002simple,PEDERSEN200747}
\beq\label{eq:gate_fidelity}
\mathcal{F}_{\rm gate} \equiv \frac{\big|\tr \left(\hat{\mathcal{G}}_{\rm tar}^\dagger \hat{\mathcal{G}}_{\rm QLG}\right)\big|^2+d'}{d'(d'+1)},
%\frac{1}{4}|\tr \left(\mathcal{G}_{\rm tar}^\dagger \mathcal{G}_{\rm QLG}\right)|^2
\eeq
where $d'$ is the logical-subspace dimension (e.g., $d'=2$ for a qubit) and the effective logical gate realized by the sequence of QLGs evolution is $\hat{\mathcal{G}}_{\rm QLG} = \hat{Q}^\dagger \hat{\mathbf{G}}(\vec{\theta},\vec{\gamma}) \hat{Q}$, where $\hat{Q}$ maps the physical Hilbert space to the logical basis (for details see~\cite{SM}).

\textit{Optimal pulse engineering (OPE).---}
We finally develop a numerical pulse engineering optimization method to optimally mitigate both the high-order FM errors and the TS errors under specific hardware constraints.
To this end, we replace the constant driving strength $\beta_0$ in Eq.~(\ref{eq:exp-cos}) by a time-dependent envelope
 \(\beta(\tau)\in\mathbb{R}\) used for optimal control. Over one period \(T\), we discretize \(\beta(\tau)\) into \(N_t\) slices, i.e., \(\vec{\beta}=\{\beta_{1},\ldots,\beta_{N_t}\}\), and then solve
\begin{equation}
\begin{aligned}
\min_{\vec{\beta}\in\mathbb{R}^{N_t}}\quad & \mathcal{L}\!\big(|\psi_{\rm QLG}(\vec{\beta})\rangle,\,|\psi_{\rm tar}\rangle\big)\\
\text{subject to}\quad & |\vec{\beta}-\beta_0|\le \delta,
%,\qquad n=1,\ldots,N_t,
\end{aligned}
\end{equation}
where \(\mathcal{L}(\cdot)\) is the task-specific loss (for state preparation, \(\mathcal{L}=1-|\langle\psi_{\rm tar}|\psi_{\rm QLG}(\vec{\beta})\rangle|^{2}\); for logical gates, \(\mathcal{L}=1-\mathcal{F}_{\rm gate}\)). The bound \(\delta>0\) encodes the maximum admissible amplitude feasibility set by the hardware. 
% (Optional constraints such as a slew-rate bound \(|\beta_{n+1}-\beta_{n}|\le \delta_{\rm slew}\) and an amplitude cap \(|\beta_n|\le \beta_{\max}\) can be included straightforwardly.) 
The discrete envelope \(\vec{\beta}\) is initialized as $\beta_0= 1/\omega_0$ %\(\beta_0= (1/\omega_0)^{1\times N_t}\) 
and then optimized with a gradient-based method; see more details in~\cite{SM}.

Figures~\figpanel{fig:figure3}{a} and ~\figpanel{fig:figure3}{b} illustrate the preparation of binomial, cat, and GKP codewords from the vacuum using OPE-based QLGs for different Trotter depths and control bounds. 
Fig.~\figpanel{fig:figure3}{a} shows that the infidelity $1-\mathcal{F}_{\rm gate}$ decreases rapidly as $N_t$ increases up to $2^{6}$ and eventually saturates around $10^{-6}$. 
Fig.~\figpanel{fig:figure3}{b} shows code-dependent convergence behaviors of the optimal control improved by relaxing the control bound. Specifically, optimal control reduces the infidelity from $\sim 10^{-2}$ to below $\sim 10^{-5}$ for $\delta/\beta_0 \ge 0.5$. 
These results show that high-fidelity bosonic-code preparation is feasible, with the performance governed by both the Trotter depth and the control bounds.
%(see further discussion in~\cite{SM}).  
In Fig.~\figpanel{fig:figure3}{c}, we randomly sample $10^2$ target single-qubit logical gates $\hat{\mathcal{G}}_{\rm tar} \in \mathbb{SU}(2)$, and show the distribution of the gate infidelity with a violin plot (colored region) for each bosonic code.
The three violins plots are unimodal with narrow interquartile ranges (below $10^{-1}$) and medians ranging from $\sim10^{-2}$ to $\sim 10^{-3}$, indicating that our QLG protocol is statistically stable across targets.
Moreover, as shown in Fig.~\figpanel{fig:figure3}{d}, all the three codes realize the elementary gates $\{ {\tt H}, {\tt S}, {\tt T} \}$ with gate errors $\lesssim 10^{-4}$ with optimal control,  which are comparable to the state-of-the-art performance of superconducting single-qubit gates~\cite{li2023error,PRXQuantum.5.030353}.

\textit{Discussion and Conclusion.---}
In summary, we present an analytical single-period Floquet protocol of bosonic codes with quantum lattice gates, avoiding slow adiabatic ramps over thousands of periods~\cite{gkp2024prl,LZGuo2025CP} and the global variational searches as in SNAP-based approaches~\cite{Liang2015praR,Simone2022prxq}. 
{Within this framework, we numerically demonstrate high-fidelity preparation of Haar-distributed target states and employ Haar-statistics benchmarks to assess the controllability of our single-period Floquet control via QLGs.} 
By combining QLGs with optimal pulse engineering, we achieve high-fidelity preparation of binomial, cat, and GKP codewords and universal logical single-qubit gates. 

{In the End Matter, we further compare our method with an adiabatic-ramping-based protocol in terms of timescale, fidelity, and noise robustness. Our single-period method significantly outperforms this adiabatic protocol, yielding improvements of up to three orders of magnitude across all considered metrics.
We also present a numerical controllability benchmark comparing the QLG and SNAP protocols. This benchmark shows that the computational resources required by our method scale linearly with the Hilbert-space dimension, and that the QLG ansatz achieves the target accuracy with fewer resources and substantially faster optimization convergence than the SNAP ansatz.}
% whereas the SNAP-based circuit exhibits a quadratic scaling, $\mathcal{O}(d^2)$

The impact of coherent noise and decoherence effect is analyzed in the the supplemental material~\cite{SM}, where the short, single-period operation times are shown to enable substantial robustness. 
{Our approach differs fundamentally, in both methodology and experimental implementation, from existing approaches~\cite{Liang2015praR,timo2020prl}, providing a hardware-compatible and efficient route to fast and high-fidelity control of CV systems and bosonic states.} Moreover, our framework can be naturally extended to two-qubit logical gates encoded on bosonic codes by designing two-mode noncommutative Floquet engineering. Our approach can also be applied to applications such as quantum communication~\cite{pirandola2020advances,wang2026highly}, quantum reservoir computing~\cite{dudas2023quantum} and quantum $t$-designs~\cite{PhysRevX.14.011013,dy4m-gq5c}.

\textit{Code availability.---} 
The computation code for producing the results in this work is available upon reasonable request.

\textit{Acknowledgments.---}
T.H and L.D. acknowledge insightful discussions with R.W. 
This work was supported by the National Natural Science Foundation of China (Grant No. 12475025). 
T.H acknowledges financial support by the Knut and Alice Wallenberg through the Wallenberg Center for Quantum Technology (WACQT). L.D. acknowledges financial support by the Knut och Alice Wallenberg stiftelse through project grant no. 2022.0090.

\newpage
% %\appendix
% \section{End Matter}

\section*{End Matter}\label{EndMatter}
\renewcommand{\thesection}{EM\arabic{section}} % Optional: custom numbering
 
\textit{Beyond adiabaticity.}---Previous works prepared bosonic code states using Hamiltonian Floquet engineering (HFE) combined with adiabatic ramping (AR)~\cite{LZGuo2025CP,gkp2024prl}. 
In those protocols, periodic driving is used to mimic an effective Hamiltonian whose eigenstates encode the logical code states, and a slowly varying ramp then prepares the target state (i.e., an arbitrary superposition of the logical code states) from a suitably chosen initial state (with nonzero overlap with the target state) via adiabatic evolution. 
However, it has been shown that AR protocols typically require evolution over \emph{thousands} of driving periods to achieve high fidelity~\cite{LZGuo2025CP,gkp2024prl}. 
Even with machine-learning-assisted optimization~\cite{guo2025RL}, the protocol remains slow and is therefore vulnerable to decoherence and control noise on current hardware.

{Our single-period Floquet (SF) approach goes beyond this paradigm by abandoning AR altogether. 
Instead, it directly synthesizes a target unitary that prepares the desired state from an arbitrary initial state within a single Floquet period. 
As a result, the SF method dramatically reduces the control time and exhibits substantially improved robustness under realistic noise conditions~\cite{SM}. 
When further combined with optimal pulse engineering, it achieves even higher accuracy without sacrificing its short-duration advantage.}

{In Table.~\ref{table:complexity}, we benchmark the performance of the SF and AR methods, as well as their optimized variants, in terms of timescale, state-preparation fidelity, and robustness under identical noisy environments. Here, robustness is defined as the maximum tolerable noise strength for achieving high-fidelity ($>0.99$) code-state preparation under the same noise model (see the supplemental material~\cite{SM} for detailed numerical setup and the full robustness analysis). 
Overall, our SF method significantly outperforms the AR protocol, yielding improvements of up to three orders of magnitude across all considered metrics.
}

\begin{table}[h!]
\centering
\renewcommand{\arraystretch}{1.2}
\begin{tabular}{c @{\hspace{0.5cm}} c @{\hspace{0.2cm}} c @{\hspace{0.2cm}} c}
    \hline \hline
    \textbf{Methods} 
    & Time & Infidelity & Robustness \\
    \hline  
    AR \cite{LZGuo2025CP,gkp2024prl} & $ \sim 10^{3}  $ & $10^{-2}$& $10^{-3}$ \\
    AR+ML \cite{guo2025RL} & $\sim 10^{2} $ &  $10^{-2}$& N/A \\
    SF (this work) & $\mathcal{O}(1) $ & $10^{-2}$&$10^{-2}$ \\
    SF+OPE (this work) & $\mathcal{O}(1)$ & $10^{-5}$&$ 10^{-1}$ \\
    \hline\hline
\end{tabular}
\caption{{Benchmark comparison of our single-period Floquet (SF) method with adiabatic ramping (AR) approach for bosonic code-state preparation. 
We compare the required preparation time (in units of the Floquet period), the achieved infidelity, and the robustness under identical noise conditions. Here, the robustness is defined as the maximum tolerable noise strength for maintaining high-fidelity ($>0.99$) code-state preparation.
Machine-learning-assisted AR (AR+ML) method and SF method with optimal pulse engineering (SF+OPE) are also included in the comparison. 
}}
\label{table:complexity}
\end{table}

\begin{figure}[h]
\centering
\includegraphics[width=\columnwidth]{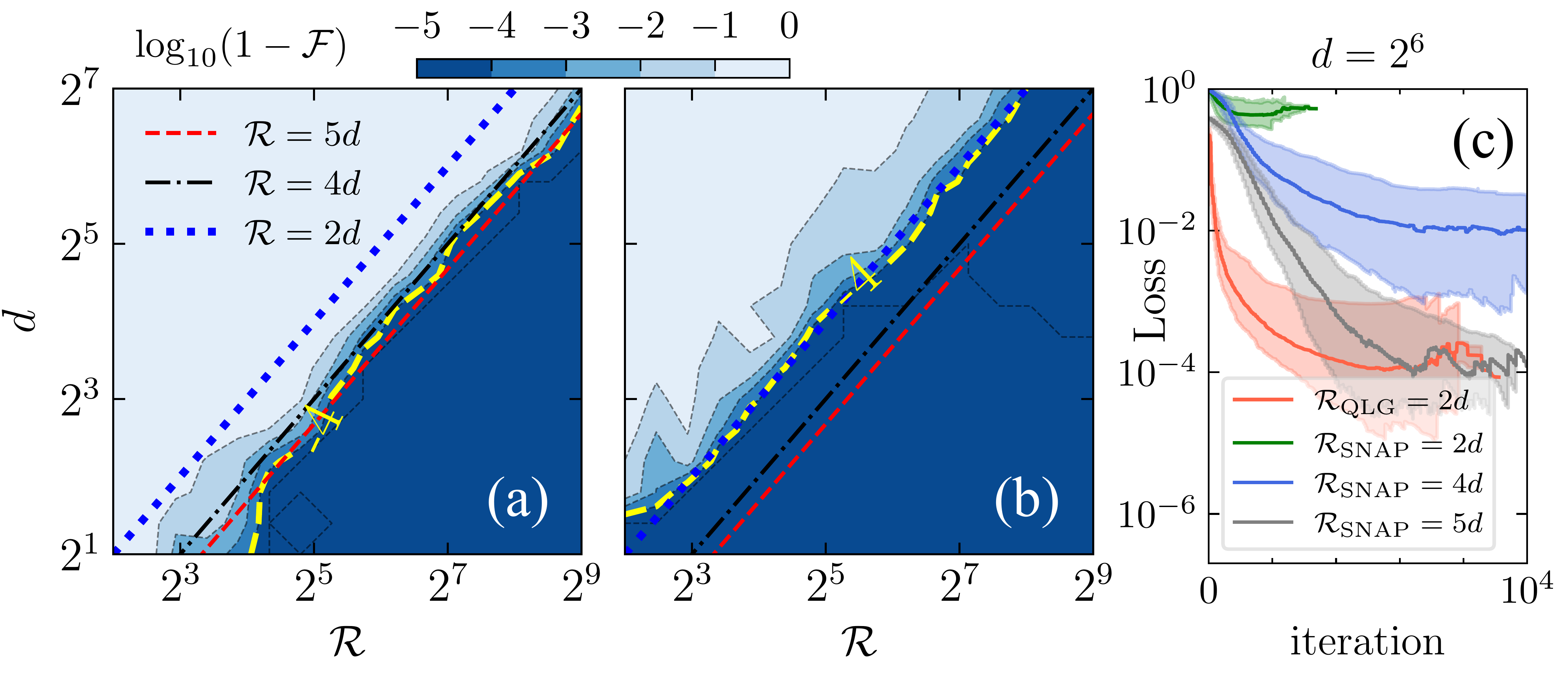}%width=\columnwidth scale=0.3
\caption{\label{fig:controllability}{
\textbf{Controllability benchmark.} Average infidelity for preparing \(10^2\) Haar-random target states as a function of the Hilbert-space dimension \(d\) and the number of trainable parameters \(\mathcal{R}\). Results for the SNAP- and QLG-based circuits are shown in panels (a) and (b), respectively. The thick yellow dashed curve marks the contour \(\mathrm{log}_{10}(1-\mathcal{F})\simeq 10^{-4}\), while the other dotted/dashed guide lines indicate linear scalings \(\mathcal{R}=2d, 4d, 5d\). (c) Training dynamics for $d=2^6$, showing the loss versus iteration under the same optimization setup for several parameter budgets. Parameters: $\delta=2$ and $\alpha_{\rm max}=\sqrt{d}$.
}}
\end{figure}

\textit{Controllability Benchmark.}---Universal control of a CV logical qubit can be realized by combining displacement operations with selective number-dependent arbitrary phase (SNAP) gates in a resonator--qubit architecture~\cite{Liang2015praR,heeres2017snap}. 
Since these operations act primarily on the resonator and the ancillary qubit, respectively, they are typically applied sequentially~\cite{Liang2015praR}. 
This gate set provides a standard building block for arbitrary CV control in truncated Hilbert spaces, with full $\mathbb{U}(d)$ controllability demonstrated experimentally in circuit QED~\cite{heeres2017snap}. 
We therefore benchmark the QLG circuit and the following $L$-layer SNAP circuit:
\begin{eqnarray}
\hat{\mathcal{C}}_{\rm snap} &=& \prod_{\ell=1}^{L}\prod_{j=1}^{d}\hat{\mathcal{D}}(\alpha_{\ell})\,\hat{S}(\vartheta_{j,\ell}),\\
\hat{\mathcal{C}}_{\rm qlg} &=& \prod_{n=1}^{N_t}{\hat{\mathbf{G}}}[\vec{\beta}(t_n)].
\end{eqnarray}
Here, $\hat{\mathcal{D}}(\alpha)$ denotes the displacement operator and $\hat{S}(\vartheta)$ is the SNAP gate. 
Since $\alpha\in\mathbb{C}$ is complex and bounded by $|\alpha|\le\alpha_{\rm max}$, each displacement contributes two real parameters (its real and imaginary parts). 
Therefore, the SNAP circuit contains $L\times (d+2)$ trainable real parameters, whereas the QLG circuit contains $N_t$ trainable parameters.

To quantify controllability under finite circuit complexity, we benchmark the preparation of Haar-random target states in a truncated Hilbert space of dimension $d$. 
Here $\mathcal{R}$ denotes the number of trainable parameters, and we define the computational resources for optimization as the minimal number of parameters required to reach a target infidelity threshold $\varepsilon$,
\begin{equation}
\mathcal{R}_{\min}(d;\varepsilon)
:=
\min\left\{
\mathcal{R}\; \middle|\;
1-\mathcal{F}^\star_{\mathcal{C}}(\mathcal{R},d)\le \varepsilon
\right\},
\end{equation}
where $\mathcal{F}^\star_{\mathcal{C}}(\mathcal{R},d)$ is the optimized fidelity achievable by a circuit ansatz $\mathcal{C}$ with $\mathcal{R}$ trainable parameters. 
In our numerics, the $\mathcal{R}$ trainable parameters are assigned by including all parameters from the first several complete SNAP layers, while any remaining parameters, if needed, are randomly sampled from the subsequent layer. 
Further details of the optimization procedure are provided in the SM~\cite{SM}.

{Figure~\ref{fig:controllability} shows the average optimized infidelity as a function of $d$ and $\mathcal{R}$ for the SNAP- and QLG-based circuits. 
The yellow dashed contour marks the threshold $\varepsilon \simeq 10^{-4}$ and therefore directly reveals the scaling of the minimal resource cost.

% For the QLG circuit, the number of trainable parameters is $\mathcal{R}_{\rm qlg}= N_t$, where $N_t$ is the Trotter number. 

{As shown in Figs.~\figpanel{fig:controllability}{a} and \figpanel{fig:controllability}{b}, the fixed-infidelity contour (yellow dashed) exhibits qualitatively different scaling behaviors for the two circuit ansatze:
\begin{equation}
\mathcal{R}^{\rm qlg}_{\min}(d;10^{-4})\sim 2d,\qquad
\mathcal{R}^{\rm snap}_{\min}(d;10^{-4})\sim 5d.
\end{equation}
Thus, to achieve the same target accuracy, the QLG ansatz requires substantially fewer trainable parameters as the Hilbert-space dimension increases.
We attribute this favorable scaling to the analytical construction underlying the QLG method, which provides an accurate closed-form initialization and thereby reduces the subsequent optimization burden.
This interpretation is supported by the close agreement between the fidelities of the bare and optimized QLG protocols, as shown in Figs.~\figpanel{fig:figure2}{a} and \figpanel{fig:controllability}{b}. This is further consistent with the training dynamics in Fig.~\figpanel{fig:controllability}{c}, where the QLG protocol converges more efficiently under the same optimization setup~\cite{SM}.

\bibliographystyle{apsrev4-1}
\bibliography{reference}

\pagebreak
\widetext
\begin{center}
\textbf{ \large Supplemental Material of \\ ``Single-Period Floquet Control of Bosonic Codes with Quantum Lattice Gates''}
\end{center}
%\appendix
\onecolumngrid

\section{Appendix A: Noncommutative Fourier transformation}
Given an arbitrary target Hamiltonian, which, in the Fock basis $\{|n\rangle\}$, can be expressed as  
\begin{equation}
\hat{H}_{\rm tar} =
%\equiv E_{\rm tar}|\psi_{\rm tar}\rangle\langle\psi_{\rm tar}|=
\sum_{n,m}h_{n,m}|n\rangle\langle m|,
\label{SMHtar}
\end{equation}
%where $E_{\rm tar}$ is an engineered energy gap that is determined by the driving potential, and 
where $h_{n,m}$ is the superposition coefficient determined by the target state. The Husimi Q-function of the target Hamiltonian can be readily obtained as
\begin{eqnarray}
H^{Q}_{\rm tar}(x,p)&=&\langle\alpha|\hat{H}_{\rm tar}|\alpha\rangle
=e^{-|\alpha|^{2}}\sum_{n,m}\sum_{n',m'}h_{n,m}\frac{(\alpha^{*})^{n'}\alpha^{m'}}{\sqrt{n'!m'!}}\langle n'|n\rangle\langle m|m'\rangle \nonumber\\
&=&e^{-|\alpha|^{2}}\sum_{n,m}h_{n,m}\frac{(\alpha^{*})^{n}\alpha^{m}}{\sqrt{n!m!}} \nonumber\\
&=&e^{-\frac{x^{2}+p^{2}}{2\lambda}}\sum_{n,m}h_{n,m}\frac{\left(\frac{x-ip}{\sqrt{2\lambda}}\right)^{n}\left(\frac{x+ip}{\sqrt{2\lambda}}\right)^{m}}{\sqrt{n!m!}},
\label{HusimiQ}
\end{eqnarray}
where $|\alpha\rangle$ denotes the coherent state defined by $\hat{a}|\alpha\rangle=\alpha|\alpha\rangle$, with $\alpha=\left(x+ip\right)/\sqrt{2\lambda}$ and $\lambda$ the dimensionless Planck constant.
By switching to polar coordinates with $x=r\cos{\theta}$ and $p=r\sin{\theta}$, and introducing $k_{x}=k\cos{\tau}=k\cos{\omega_{0}t}$ and $k_{p}=k\sin{\tau}=k\sin{\omega_{0}t}$, where $\omega_{0}$ is the resonance frequency of the cavity, the noncommutative Fourier coefficient corresponding to the target Hamiltonian [cf. Eq.~(3) in the main text] can be expressed as
\begin{equation}
f_{\rm tar}(k,\tau)=\frac{e^{\frac{\lambda}{4}k^{2}}}{2\pi}\int\int rdrd\theta H^{Q}_{\rm tar}(r\cos{\theta},r\sin{\theta})e^{-ikr\cos{(\theta-\tau)}}.
\label{fTdef}
\end{equation} 
Under the Fock basis, each component of $f_{\rm tar}$ (i.e., each matrix element) can be derived as
\begin{eqnarray}
f_{n,m}(k,\tau)&=& \frac{\langle n|f_{\rm tar}(k,\tau)|m\rangle}{h_{n,m}} \nonumber\\
&=&\ \frac{e^{\frac{\lambda}{4}k^{2}}}{2\pi h_{n,m}}}\int_{0}^{+\infty}rdr\int_{0}^{2\pi}d\theta\langle n|H^{Q}_{\rm tar}(r\cos{\theta},r\sin{\theta})|m\rangle e^{-ikr\cos{(\theta-\tau)} \nonumber\\
&=&\frac{e^{\frac{\lambda}{4}k^{2}}}{2\pi}\frac{e^{i(m-n)\tau}}{\sqrt{n!m!}}\left(\frac{1}{\sqrt{2\lambda}}\right)^{n+m}\int_{0}^{+\infty}r^{n+m+1}e^{-\frac{r^{2}}{2\lambda}}dr\int_{0}^{2\pi}e^{-ikr\cos{(\theta-\tau)}}e^{i(m-n)(\theta-\tau)}d\theta \nonumber\\
&=&e^{\frac{\lambda}{4}k^{2}}\frac{e^{i(m-n)\tau}}{\sqrt{n!m!}}i^{n-m}\left(\frac{1}{\sqrt{2\lambda}}\right)^{n+m}\int_{0}^{+\infty}r^{n+m+1}e^{-\frac{r^{2}}{2\lambda}}J_{n-m}(-kr)dr \nonumber\\
&=&\left \{ \begin{array}{lll}
e^{\frac{\lambda}{4}k^{2}}\sqrt{\frac{n!}{m!}}\left(-ie^{i\tau}\frac{1}{k}\sqrt{\frac{2}{\lambda}}\right)^{m-n}\frac{\lambda}{\Gamma(1+n-m)}{}_{1}F_{1}(1+n; 1+n-m; -\frac{\lambda}{2}k^{2}), \quad n-m>-1, \\
e^{\frac{\lambda}{4}k^{2}}\sqrt{\frac{m!}{n!}}\left(-ie^{-i\tau}\frac{1}{k}\sqrt{\frac{2}{\lambda}}\right)^{n-m}\frac{\lambda}{\Gamma(1+m-n)}{}_{1}F_{1}(1+m; 1+m-n; -\frac{\lambda}{2}k^{2}), \quad n-m\leq -1.
\end{array} \right.
\label{fTcomponent}
\end{eqnarray} 
Here, $J_{n-m}(z)$ is the Bessel function of the first kind (of order $n-m$), which satisfies the identity
\begin{equation}
J_{z}(x)=\frac{i^{-z}}{2\pi}\int_{0}^{2\pi}e^{i(zy+x\cos{y})}dy,
\end{equation}
$\Gamma(n)=(n-1)!$ is the Gamma function, and ${}_{1}F_{1}(a;b;z)$ is the Kummer confluent hypergeometric function. Note that, in the final step of Eq.~(\ref{fTcomponent}), we have classified the identity into two different situations for the convenience of numerical simulations. The entire noncommutative Fourier coefficient can be reconstructed as
\begin{equation}
f_{\rm tar}(k,\tau)=\sum_{n,m}h_{n,m}f_{n,m}(k,\tau).
\label{fTcomplete}
\end{equation}
The corresponding driving potential can then be obtained as~\cite{guo2024prl,LZGuo2025CP}
\begin{eqnarray}
\hat{V}(\hat{x},\tau) &=& \int_{-\infty}^{+\infty} \frac{|k|}{2} \beta_0 f_{\rm tar}(k, \tau) e^{ik\hat{x}} \, dk \nonumber\\
&=&\int_{0}^{+\infty}\beta_0 A(k,\tau)\cos{[k\hat{x}+\phi(k,\tau)]}dk,
\end{eqnarray}
where $\beta_0$ is a \emph{target-independent} parameter that controls the driving strength. The time-dependent amplitude $A(k,\tau)$ and phase $\phi(k,\tau)$ are given by
\begin{equation}
A(k,\tau)=k|f_{\rm tar}(k\cos{\tau},k\sin{\tau})|,\quad \phi(k,\tau)=\text{Arg}[f_{\rm tar}(k\cos{\tau},k\sin{\tau})].
\end{equation}

\section{Appendix B: Floquet Hamiltonian engineering}
To generate the target Hamiltonian $\hat{H}_{\rm tar}$, which is in general an arbitrary function of quadrature operators $\hat{x}$ and $\hat{p}$, we drive the cavity with a periodic external potential $\hat{V}(\hat{x},t)=V(\hat{x},t+T_d)$ with $T_d=2\pi/\omega_d$ and $\omega_d$ the driving frequency. The total Hamiltonian of the driven cavity is given by
\begin{equation}
\hat{H}_{\rm cav}(t)=\frac{\omega_0}{2}\left(\hat{x}^{2}+\hat{p}^{2}\right)+ \hat{V}(\hat{x},t).
\label{rampH}
\end{equation}
To analyze the system dynamics, we transform the above Hamiltonian into a rotating frame at frequency $\Omega=2\pi/T$ with $T=nT_d $ $(n\in \mathbb{Z}^+)$. The quadrature operator $\hat{x}$ transforms under the unitary operator $\hat{O}(t)\equiv e^{i\hat{a}^\dagger\hat{a}\Omega t}$ as $\hat{O}(t)\hat{x}\hat{O}^\dagger(t)=\hat{x}\cos (\Omega t)+\hat{p}\sin (\Omega t)$, while the transformed Hamiltonian in the rotating frame becomes
\beqa\label{eq-Ht}
\hat{H}(t)&\equiv&\hat{O}(t)\hat{\mathcal{H}}(t)\hat{O}^\dagger(t)-i\lambda \hat{O}(t)\dot{\hat{O}}^\dagger(t)\nonumber \\
&=&\beta V\Big[\hat{x}\cos (\Omega t)+\hat{p}\sin (\Omega t),t\Big].
\eeqa
Here,  we have employed the multiphoton resonance condition $T=2\pi/\omega_0$, or equivalently $\Omega=\omega_0$, such that the driving frequency is an integer multiple of the bare cavity frequency.

A system subject to a periodic drive is commonly referred to as a \textit{Floquet system}~\cite{Shirley1965pr}. The Floquet theory states that the stroboscopic time evolution of such a system can be described by an effective time-independent Floquet Hamiltonian $\hat{H}_{\text F}$, which is defined by the relation~\cite{Shirley1965pr}
\beqa\label{eq-HFt0}
\exp\Big(-i\frac{1}{\lambda}\hat{H}_{\text F}T\Big)=\mathcal{T}\exp\Big[-i\frac{1}{\lambda}\int_{0}^{T}\hat{H}(t)dt\Big], 
\eeqa
where $\mathcal{T}$ is the time-ordering operator. 
Under the rotating wave approximation (RWA), the Floquet Hamiltonian $\hat{H}_F$ can be well approximated by the time average of $\hat{H}(t)$ over one driving period $T$ ~\cite{guo2024prl,Eckardt2015NJP}, i.e.,
\beqa\label{eq-h0h1h3}
\lim_{\omega_0/\beta\rightarrow\infty}\hat{H}^{}_{\text F}(\hat{x},\hat{p})&=&\frac{1}{T}\int_{0}^{T}dt  \hat{H}(t).\ \ \ 
\eeqa
By properly engineering the time-dependent driving potential $\hat{V}(\hat{x},t)$, one can realize a desired Floquet Hamiltonian $\hat{H}_{\text F}(\hat{x},\hat{p})$ that effectively implements the target Hamiltonian $\hat{H}_{\rm tar}(\hat{x},\hat{p})$.

\section{Appendix C:. Analytic Constructions in Truncated Fock Spaces: Principal-Branch Generators and Householder Unitary Synthesis}

In this section, we present two core analytical derivations: (i) the construction of the principal-branch Hamiltonian that generates a given unitary $\hat{U}\in \mathbb{U}(N)$ on a truncated Fock space, and (ii) a closed-form synthesis of a unitary operator that maps a prescribed normalized initial state to a normalized final state, achieved via a Householder reflection combined with a rank-one phase correction. 
%Both derivations are self-contained and include concise algebraic verifications.

\subsection{Appendix C.1: Principal-branch Hamiltonian from a given unitary}
\label{sec:principal-branch}

Let $\hat{U}\in \mathbb{U}(N)$ be a unitary operator acting on the truncated Fock space
$\mathrm{span}\{\ket{0},\ldots,\ket{N-1}\}$ and assume
that $\hat{U}$ is implemented as the time evolution under a (priori unknown)
time–independent Hamiltonian $\hat{H}$ over a finite duration $T>0$, i.e.,
\begin{equation}
    \hat{U} = \exp\!\left(-\frac{\mathrm{i}}{\hbar} \hat{H}T\right),\qquad T>0 .
    \tag{B1}
    \label{B1}
\end{equation}
Our goal is to reconstruct a Hermitian generator
$\hat{H}_{\mathrm{pr}}$ such that: (i) it reproduces the given unitary $\hat{U}$, and
(ii) its eigenvalues lie on the principal branch of the complex logarithm.
We refer to such a generator $\hat{H}_{\mathrm{pr}}$ as the \emph{principal-branch
Hamiltonian} associated with $\hat{U}$.

By the spectral theory, there exists a unitary $\hat{V}$ and eigenvalues
$\lambda_k$ of $\hat{U}$ such that
\begin{equation}
    \hat{U} = \hat{V} \, \mathrm{diag}(\lambda_1,\ldots,\lambda_N)\, \hat{V}^\dagger .
    \tag{B2}
\end{equation}
Since $\hat{U}$ is unitary, each eigenvalue can be written as $\lambda_k = \mathrm{e}^{-\mathrm{i}\theta_k}$
with some real phases $\theta_k$. To obtain a unique logarithm, we choose the
\emph{principal arguments} $\theta_k\in(-\pi,\pi]$, i.e., we restrict
all eigenphases to this interval. This choice amounts to using the
principal branch of the complex logarithm, which resolves the ambiguity in the
generator up to integer multiples of $2\pi/T$.

The principal matrix logarithm of $\hat{U}$ is then defined via the holomorphic functional calculus as
\begin{equation}
    \log \hat{U}
    \equiv \hat{V} \,\mathrm{diag}\bigl(\log\lambda_1,\ldots,\log\lambda_N\bigr)\, \hat{V}^\dagger
     = -\mathrm{i}\, \hat{V}\,\mathrm{diag}(\theta_1,\ldots,\theta_N)\,\hat{V}^\dagger .
    \tag{B3}
\end{equation}
Here $\log\lambda_k$ denotes the complex logarithm evaluated on the principal branch, i.e., $\log\lambda_k=-\mathrm{i}\theta_k$ with
$\theta_k\in(-\pi,\pi]$.

Taking the principal logarithm of Eq.~(\ref{B1}) yields
\begin{equation}
    -\frac{\mathrm{i}}{\hbar}\hat{H}T = \log \hat{U}
    \qquad\Longrightarrow\qquad
    \hat{H}_{\mathrm{pr}} = \frac{\mathrm{i}\hbar}{T}\,\log \hat{U}
                    = \frac{\hbar}{T}\,
                      \hat{V}\,\mathrm{diag}(\theta_1,\ldots,\theta_N)\,\hat{V}^\dagger .
    \tag{B4}
    \label{B4}
\end{equation}
Thus, $\hat{H}_{\mathrm{pr}}$ is Hermitian and generates $\hat{U}$ over time $T$:
$\exp(-\frac{\mathrm{i}}{\hbar}\hat{H}_{\mathrm{pr}}T)=\hat{U}$, with all its eigenvalues $\frac{\hbar}{T}\theta_k$ lying within the interval
$(-\pi\hbar/T,\pi\hbar/T]$. In particular, $\hat{H}_{\mathrm{pr}}$ is the
\emph{unique} such generator, as long as none of the eigenvalues of $\hat{U}$ lies exactly at
$-1$ (corresponding to $\theta_k=\pm\pi$), which would place them on the branch
cut of the complex logarithm.

Equivalently, let $\ket{u_k}$ be an orthonormal eigenbasis of $\hat{U}$, satisfying $\hat{U}\ket{u_k}=\lambda_k\ket{u_k}$, and define the rank-one spectral projectors
\begin{equation*}
    \hat{P}_k := \ket{u_k}\!\bra{u_k} .
\end{equation*}
Then Eq.~(\ref{B4}) can be expressed in the compact spectral form
\begin{equation}
    \hat{H}_{\mathrm{pr}}
      = \frac{\hbar}{T}\sum_{k=1}^{N}\theta_k \hat{P}_k,
    \tag{B5}
\end{equation}
which explicitly shows that $\hat{H}_{\mathrm{pr}}$ acts by multiplying each
eigenvector $\ket{u_k}$ of $\hat{U}$ by the corresponding eigenvalue
$\frac{\hbar}{T}\theta_k$.

In the Fock basis $\{\ket{n}\}_{n=0}^{N-1}$, the matrix elements of $\hat{H}_{\rm pr}$ are given by
\begin{equation}
    H_{mn}
      = \left[\frac{\hbar}{T}
          \hat{V}\,\mathrm{diag}(\theta_k)\,\hat{V}^\dagger
        \right]_{mn}
      = \frac{\hbar}{T}\sum_{k=1}^{N}\theta_k
        \braket{m|u_k}\braket{u_k|n},
    \tag{B6}
    \label{B6}
\end{equation}
which provides a practical recipe for computing $\hat{H}_{\mathrm{pr}}$ from a
numerically known unitary $\hat{U}$: diagonalize $\hat{U}$, extract the principal
arguments of its eigenvalues, and construct $\hat{H}_{\mathrm{pr}}$ via
Eq.~(\ref{B6}). This procedure is employed in the main text to associate an
effective Hamiltonian with the synthesized unitary gates acting on the truncated space of a harmonic oscillator.

\subsection{Appendix C.2: Unitary synthesis via Householder transformation}
\label{sec:householder}

In this section, we present an explicit construction of a unitary operator that
maps a given pure state $\ket{\psi_i}$ to another pure state
$\ket{\psi_f}$ within the same Hilbert space $\mathcal{H}_N$. The
construction is based on the (quantum) Householder transformation~\cite{Urias2010}, which corresponds to a reflection across a hyperplane in $\mathcal{H}_N$.
Throughout this section, we assume that both states are normalized, i.e., $\braket{\psi_i|\psi_i} = \braket{\psi_f|\psi_f}=1$.

We begin by introducing the complex overlap between the initial and final states:
\begin{equation}
r \equiv \braket{\psi_f | \psi_i} \in \mathbb{C},
\end{equation}
and define a phase factor $\mathrm{e}^{\mathrm{i}\phi}$ such that
\begin{equation}
    \mathrm{e}^{\mathrm{i}\phi} =
    \begin{cases}
        \dfrac{r}{|r|}, & r \neq 0, \\
        1,                        & r = 0.
    \end{cases}
    \label{eq:phi_def}
    %\tag{B7}
\end{equation}
It is clear that $\mathrm{e}^{-\mathrm{i}\phi}r= |r|\ge 0$ is real and
nonnegative: geometrically, this operation removes only the global phase of the overlap and does not affect the relative orientation of the two states in Hilbert space.

\paragraph{Generic case: non–collinear states.}
Suppose that $\ket{\psi_i}$ and $\ket{\psi_f}$ are not collinear, i.e.\ $|r| \neq 1$. We define the vector
\begin{equation}
    \ket{v} \equiv \ket{\psi_i} - \mathrm{e}^{\mathrm{i}\phi}\ket{\psi_f},
    \qquad
    \ket{u} \equiv \frac{\ket{v}}{\|\ket{v}\|},
\end{equation}
and introduce the corresponding Householder reflector~\cite{Urias2010}%atkinson2008introduction
\begin{equation}
    \hat{H}_{h} \equiv \hat{I} - 2 \ket{u}\!\bra{u}.
    \label{eq:H_def}
    %\tag{B8}
\end{equation}
By construction, $\hat{H}_{h}$ is Hermitian (i.e., $\hat{H}_{h}^\dagger = \hat{H}_{h}$) and satisfies
\begin{equation}
    \hat{H}_{h}^2 = \left(\hat{I} - 2\ket{u}\!\bra{u} \right)^2
        = \hat{I} - 4\ket{u}\!\bra{u} + 4\ket{u}\!\braket{u|u}\!\bra{u}
        = \hat{I},
\end{equation}
which shows that $\hat{H}_{h}$ is also unitary. Thus, $\hat{H}_{h}$ represents a reflection about the hyperplane orthogonal to $\ket{u}$.

Next, we introduce a phase gate that acts nontrivially only on the target state:
\begin{equation}
    \hat{P}_{h} \equiv \hat{I} + \left(\mathrm{e}^{-\mathrm{i}\phi}-1\right)\ket{\psi_f}\!\bra{\psi_f},
    \label{eq:P_def}
    %\tag{B9}
\end{equation}
which multiplies $\ket{\psi_f}$ by $\mathrm{e}^{-\mathrm{i}\phi}$ and
acts as the identity on any state orthogonal to $\ket{\psi_f}$.

Finally, we construct the unitary operator
\begin{equation}
    \hat{U} \equiv \hat{P}_{h} \hat{H}_{h}.
    \label{eq:U_def}
    %\tag{B10}
\end{equation}
Since both $\hat{P}_{h}$ and $\hat{H}_{h}$ are unitary, their product $\hat{U}$ is also unitary. We now verify explicitly that
$\hat{U}$ maps $\ket{\psi_i}$ to $\ket{\psi_f}$ as desired.

\textit{Proof.}
Using the identity $\mathrm{e}^{-\mathrm{i}\phi}r = |r|$, we first compute the norm of $\ket{v}$:
\begin{align}
    \| \ket{v} \|^2
      &= \braket{v|v}
       = \braket{\psi_i|\psi_i}
         - \mathrm{e}^{-\mathrm{i}\phi}\braket{\psi_f|\psi_i}
         - \mathrm{e}^{\mathrm{i}\phi}\braket{\psi_i|\psi_f}
         + \braket{\psi_f|\psi_f} \notag \\
      &= 2 - \mathrm{e}^{-\mathrm{i}\phi}r
              - \mathrm{e}^{\mathrm{i}\phi}r^*
       = 2\bigl(1-|r|\bigr).
    \label{eq:v_norm}
    %\tag{B11}
\end{align}
We also have
\begin{equation}
    \braket{v|\psi_i}
      = 1 - \mathrm{e}^{-\mathrm{i}\phi}r
      = 1 - |r|.
\end{equation}
For $|r|<1$, this quantity is nonzero, and one can evaluate the action of $\hat{H}_{h}$:
\begin{align}
    \hat{H}_{h}\ket{\psi_i}
      &= \ket{\psi_i}
         - 2\ket{v}\frac{\braket{v|\psi_i}}{\| \ket{v}\|^2} \notag \\
      &= \ket{\psi_i}
         - 2\ket{v}\,
           \frac{1-|\alpha|}{2(1-|\alpha|)}
       = \ket{\psi_i} - \ket{v}
       = \mathrm{e}^{\mathrm{i}\phi}\ket{\psi_f}.
    \label{eq:H_on_psi_i}
    %\tag{B12}
\end{align}
Therefore, the Householder reflector $\hat{H}_{h}$ maps the initial state to the target state up to the controlled phase factor $\mathrm{e}^{\mathrm{i}\phi}$.

By construction of $\hat{P}_{h}$, we have
\begin{equation}
    \hat{P}_{h}\ket{\psi_f} = \mathrm{e}^{-\mathrm{i}\phi}\ket{\psi_f},
\end{equation}
and $\hat{P}_{h}$ acts as the identity on any vector orthogonal to $\ket{\psi_f}$.
Combining this with Eq.~\eqref{eq:H_on_psi_i}, we finally obtain
\begin{equation}
    \hat{U}\ket{\psi_i}
      = \hat{P}_{h} \hat{H}_{h} \ket{\psi_i}
      = \hat{P}_{h}\bigl(\mathrm{e}^{\mathrm{i}\phi}\ket{\psi_f}\bigr)
      = \ket{\psi_f}.
    \label{eq:U_maps}
    %\tag{B13}
\end{equation}
Hence, $\hat{U}$ is a desired unitary operator that exactly transforms the initial
state $\ket{\psi_{i}}$ into the target state $\ket{\psi_{f}}$ whenever $|\alpha|\neq 1$.

\paragraph{Collinear case.}
If the two states differ only by a global phase, i.e., $\ket{\psi_f} = \mathrm{e}^{\mathrm{i}\theta}\ket{\psi_i}$ with
$\theta\in\mathbb{R}$, the above construction becomes singular since
$\ket{v}=0$. In this special case, we can directly define the unitary operator as
\begin{equation}
    \hat{U} = \hat{I} + (\mathrm{e}^{\mathrm{i}\theta}-1)\ket{\psi_i}\!\bra{\psi_i},
    \label{eq:collinear-U}
    %\tag{B14}
\end{equation}
which is simply a one-dimensional phase operation.  It is straightforward to verify that $\hat{U}$ is unitary and satisfies
$\hat{U} \ket{\psi_i} = \mathrm{e}^{\mathrm{i}\theta}\ket{\psi_i}
 = \ket{\psi_f}$.

\medskip
To summarize, Eqs.~\eqref{eq:phi_def}--\eqref{eq:collinear-U} provide an
explicit and constructive method for synthesizing a unitary operator that maps an arbitrary
normalized initial state $\ket{\psi_i}$ to a normalized target state $\ket{\psi_f}$.
% In the generic case where the states are not collinear, the unitary is realized as the product of a Householder reflection
% and a single-qubit phase gate. In the special case where the states differ only by a global phase, the unitary reduces to a simple one-dimensional phase rotation on $\ket{\psi_i}$. This construction is employed in the main text to implement state–to–state transformations and, more generally, to construct multi–state unitaries.
\\

{Note that this construction strategy is not unique. For two normalized states $|\psi_i\rangle$ and $|\psi_f\rangle$, there exist infinitely many unitary operators $\hat{U}$ satisfying $\hat{U}|\psi_i\rangle = |\psi_f\rangle$. In principle, one may adopt different approaches, including Householder reflections~\cite{Urias2010}, basis-completion methods~\cite{Edelman1998}, or numerical optimization on constrained matrix manifolds~\cite{Absil2008}.
In the present work, we adopt the Householder construction because it provides a closed-form and non-iterative solution that is unitary by construction and maps the initial state to the target state exactly, up to machine precision in numerical implementation. To illustrate its numerical performance, we show in Fig.~\ref{fig:figure_Householder} the infidelity of the Householder-based unitary construction as a function of the Hilbert-space dimension, where the state fidelity is defined as ($F =|\langle\psi_f|U|\psi_i\rangle|^2$). For randomly chosen initial and target states, the mapping infidelity remains at the level of $10^{-15}$ even up to dimension $d=2^{13}$. Each error bar denotes the standard deviation over $10^2$ numerical trials, and the full simulation takes less than 10 seconds on a standard laptop. A key practical advantage of the Householder method is that it depends only on the two states of interest and does not require explicitly constructing the remaining orthogonal subspace. Moreover, unlike iterative numerical approaches, it avoids initialization dependence, convergence issues, and residual optimization errors.}

\begin{figure}[h]
\centering
\includegraphics[scale=0.8]{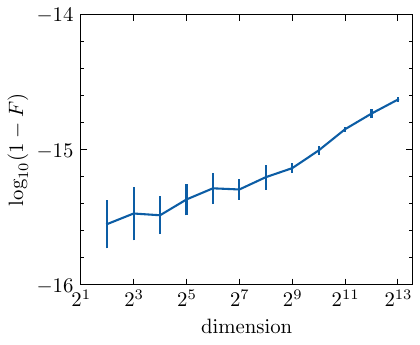} %%% width =\columnwidth scale=0.3
\caption{\label{fig:robustness}
{Logarithm of the state-mapping infidelity for the Householder-based unitary construction as a function of the Hilbert-space dimension. Error bars denote the standard deviation over $10^2$ numerical trials.
}}
\label{fig:figure_Householder}
\end{figure}

\section{Appendix D:. Haar-Random States}
In this section, we aim to elucidate the construction procedure of unitary operators for generating Haar-random states---quantum states that are uniformly sampled from the unit sphere in a $d$-dimensional complex Hilbert space $\mathbb{C}^d$, according to the unique unitarily invariant Haar measure on the unitary group $\mathbb{U}(d)$.

\medskip

\begin{figure}[h]
    \centering
    \includegraphics[width=0.6\linewidth]{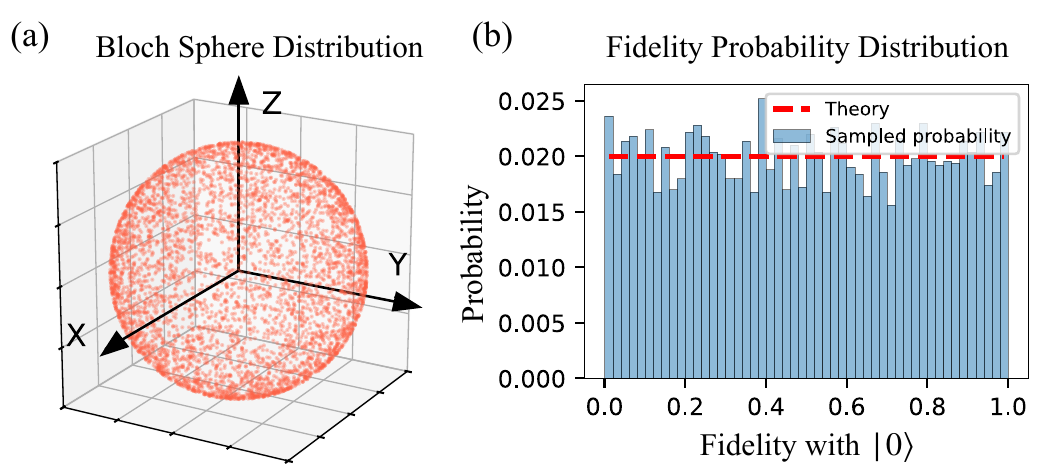}
    \caption{\textbf{Two-dimensional Haar-random states.} The distribution of \(10^4\) two-dimensional Haar-random states (\(d=2\)) is illustrated on the Bloch sphere in panel (a), and the corresponding probability distribution is shown in panel (b), in comparison with the theoretical prediction given in Eq.~\eqref{Aeq:Haar}.
    }
    \label{fig:2dHaar}
\end{figure}

Haar-random pure states play a fundamental role in quantum information theory, as they capture typical properties of quantum systems in high-dimensional Hilbert space. A pure state $\ket{\psi} \in \mathbb{C}^d$ is said to be Haar-distributed if it is sampled uniformly from the unit sphere $S^{2d-1} \subset \mathbb{C}^d$ with respect to the unitarily invariant Haar measure on the unitary group $\mathbb{U}(d)$. Mathematically, a Haar-random pure state $\ket{\psi_{\rm Haar}}$ satisfies the invariance condition $\ket{\psi_{\rm Haar}} \sim \hat{U} \ket{\psi_{\rm Haar}}$ for any unitary $\hat{U} \in \mathbb{U}(d)$.
This invariance property makes Haar-random states an essential tool in the study of quantum typicality, random quantum circuits, quantum chaos, and the benchmarking of quantum devices.

A computationally efficient method for generating such states is to normalize a complex Gaussian vector, as discussed in Refs.~\cite{zyczkowski2001induced}. Specifically, each component of a complex vector $\vec{z} \in \mathbb{C}^d$ is sampled independently from the standard complex normal distribution:
\begin{equation}
z_i = a + ib, \quad \text{where} \quad a, b \sim \mathcal{N}(0,1),
\end{equation}
where $\mathcal{N}$ denotes the normal distribution. The Haar-random pure state is then obtained by normalization:
\begin{equation}
\ket{\psi_{\rm Haar}} = \frac{\vec{z}}{|\vec{z}|}.
\end{equation}
This procedure generates pure states that are uniformly distributed on the complex unit sphere and hence invariant under the action of any unitary operator. The correctness of this method follows from the spherical symmetry of the complex Gaussian distribution and the unitarily invariant nature of the normalization operation, which together ensure exact sampling from the Haar measure.

Given a fixed pure reference state $\ket{\psi_{\rm ref}}$, the fidelity between a Haar-random state $\ket{\psi_{\rm Haar}}$ and the reference state is defined as
\beq
\mathcal{F} = |\langle \psi_{\rm ref}|\psi_{\rm Haar}\rangle|^2.
\eeq
This fidelity follows the probability density function (PDF)
\beq\label{Aeq:Haar}
\mathbb{P}_{\rm Haar}(\mathcal{F}) = (d - 1)(1 -\mathcal{F})^{d - 2}, \quad \mathcal{F} \in [0, 1],
\eeq
with the corresponding expected value
$\mathbb{E}[\mathcal{F}] = \frac{1}{d}$,
indicating that the average overlap between a Haar-random state and any fixed reference state decreases inversely with the Hilbert-space dimension $d$.

In \figref{fig:2dHaar}, we illustrate $10^4$ Haar-random pure states for $d=2$, uniformly distributed over the Bloch sphere [panel (a)], together with their fidelity distribution with respect to a fixed reference state [panel (b)], which in this case is uniform in $[0,1]$ in agreement with Eq.\eqref{Aeq:Haar}. Moreover, the average fidelity $\mathbb{E}[F] \approx 0.497$.

\section{Appendix E:. Preliminary and background of Bosonic codes}

Bosonic code states are naturally well-suited for autonomous quantum error correction (AQEC), as their encoded subspaces are intrinsically resilient to certain dominant error mechanisms in bosonic systems. Such states are engineered to be symmetric with respect to the dominant noise channels, such as photon loss and dephasing. Given a bosonic code with two logical states $|\bar{0}\rangle$ and $|\bar{1}\rangle$, the \emph{Knill–Laflamme condition}~\cite{Knill97pra,Knillooprl} guarantees that any encoded quantum state of the form $|\psi\rangle = c_0 |\bar{0}\rangle + c_1 |\bar{1}\rangle$, when subjected to a correctable set of error operators $\{\hat{\varepsilon}_i\}$, is mapped into an orthogonal error subspace without altering the encoded information in the (relative) amplitudes $c_0$ and $c_1$. Specifically, the condition requires
\begin{eqnarray}
\langle \bar{0} | \hat{\varepsilon}_k^\dagger \hat{\varepsilon}_l | \bar{0} \rangle &=& \langle \bar{1} | \hat{\varepsilon}_k^\dagger \hat{\varepsilon}_l | \bar{1} \rangle, \nonumber\\
\langle \bar{0} | \hat{\varepsilon}_k^\dagger \hat{\varepsilon}_l | \bar{1} \rangle &=& \langle \bar{1} | \hat{\varepsilon}_k^\dagger \hat{\varepsilon}_l | \bar{0} \rangle = 0.
\end{eqnarray}
% \begin{equation}
% \langle \psi_i | \hat{\varepsilon}_k^\dagger \hat{\varepsilon}_l | \psi_j \rangle = C_{kl} \delta_{ij}, \quad \forall |\psi_i\rangle, |\psi_j\rangle \in \mathcal{C},
% \end{equation}
These relations ensure that the error processes are logically indistinguishable, i.e., they act identically on different logical states. As a result, perfect error recovery is possible without perturbing the encoded quantum information. This makes bosonic codes a compelling architecture for scalable and fault-tolerant quantum computation. In the rest part of this section, we briefly introduce and discuss three types of widely studied bosonic codes, i.e., binomial, cat, and Gottesman-Kitaev-Preskill (GKP) states.

\subsection{Appendix E.1: Binomial codes}
Binomial codes constitute an important class of bosonic codes designed to protect logical qubits encoded in bosonic modes against photon loss, gain, and dephasing errors \cite{Michael16prx,hu2019binomial}. Unlike cat codes, which, as will be introduced later, are constructed from superpositions of coherent states, binomial codes rely on carefully engineered finite superpositions of Fock states, with the superposition coefficients determined by binomial distributions. This structure enables the exact correction of a predefined set of error operators up to a fixed order, in accordance with the Knill-Laflamme conditions. 

AQEC can be realized by tailoring the system-environment interactions such that the intrinsic dynamics of the system autonomously suppress errors and stabilize the logical subspace. These engineered dissipative processes drive the system back toward the code space following error events, thereby maintaining the error-correction conditions without requiring active measurements or feedback.

As a concrete example, we consider a low-order binomial code with logical states
\beqa\label{eq:binomialstates}
\left \{ \begin{array}{lll}
 |\bar{0}_L\rangle_{\rm bin} = \frac{1}{2}(|0\rangle+\sqrt{3}|4\rangle),\\
|\bar{1}_L\rangle_{\rm bin} =  \frac{1}{2}(\sqrt{3}|2\rangle+|6\rangle),
\end{array}\right.
\eeqa
where \(|n\rangle\) denote Fock states, and the corresponding coefficients \(c_n\) are chosen according to binomial weights such that the Knill-Laflamme conditions mentioned above are satisfied. Importantly, binomial codes possess discrete rotational symmetry: they are invariant under a $360^{\circ}/N$ phase-space rotation, where $N=2$ for the case in Eq.~(\ref{eq:binomialstates}). This symmetry implies that such codes can detect up to $N-1$ photon loss events.

Experimentally, binomial codes have been realized in superconducting circuit quantum electrodynamics (cQED) architectures, where they exhibit significantly enhanced logical-qubit lifetimes and enable high-fidelity gate operations~\cite{hu2019binomial}. These achievements constitute an important step toward scalable, fault-tolerant quantum computing based on bosonic modes.

\subsection{Appendix E.2: Cat codes}

The four-component (four-legged) cat code encodes a logical qubit into a superposition of four coherent states positioned at the corners of a square in phase space:
\beqa\label{eq-sm-4cat}
\left \{ \begin{array}{lll}
|\bar{0}_L\rangle_{\rm cat}\equiv\frac{1}{\sqrt{\mathcal{N}_0}}\big(|\alpha\rangle+|-\alpha\rangle+|i\alpha\rangle+|-i\alpha\rangle\big)\label{eq-4cat_c1}\\
|\bar{1}_L\rangle_{\rm cat}\equiv\frac{1}{\sqrt{\mathcal{N}_1}}\big(|\alpha\rangle+|-\alpha\rangle-|i\alpha\rangle-|-i\alpha\rangle\big) \label{eq-4cat_c2}
\end{array}\right.
\eeqa
where the normalization factors are given by $\mathcal{N}_{m}=8e^{-\alpha^2}[\cosh\alpha^2+(-1)^{m}\cos\alpha^2]$ for $m=0, 1$. This code exhibits a $\mathbb{Z}_4$ rotational symmetry in phase space, and enables protection against both photon loss and dephasing errors by encoding logical information along orthogonal quadrature axes~\cite{mirrahimi2014dynamically}.

Notably, the four-component cat code satisfies the Knill–Laflamme conditions~\cite{Knill97pra,Knillooprl} when the coherent amplitude $\alpha$ is tuned to specific ``sweet spots'' determined by the transcendental equation
\begin{equation}
\tan\alpha^2 = - \tanh\alpha^2.
\end{equation}
At these points, the code states exhibit optimal robustness against single-photon loss, as the resulting error states remain distinguishable and correctable~\cite{joshi2021quantum}. For this four-component cat code, the logical states $|\bar{0}_{L}\rangle_{\rm cat}$ and $|\bar{1}_{L}\rangle_{\rm cat}$
return to themselves (up to a trivial global amplitude factor) after undergoing four sequential single-photon losses~\cite{LZGuo2025CP}. This ensures protection over multiple photon loss events without decoding failure.

\subsection{Appendix E.3: GKP codes}

The GKP code~\cite{gkp2001pra,gkp2024prl} is a prominent example of translationally invariant bosonic codes, encoding logical qubits into the infinite-dimensional Hilbert space of a harmonic oscillator. It enables the correction of small displacement errors in both the position and momentum quadratures. Serving as a powerful interface between continuous-variable and discrete-variable quantum information processing, the GKP code plays a pivotal role in realizing fault-tolerant operations in bosonic systems.

As an example, the so-called ``square-lattice GKP code'' can be formulated as a bosonic stabilizer code, with stabilizer generators defined by displacement operators:
\beq
\hat{S}_q = e^{2i\sqrt{\pi} \hat{q}}, \quad \hat{S}_p = e^{-2i\sqrt{\pi} \hat{p}}.
\eeq
The logical code space is defined as the simultaneous $+1$ eigenspace of these commuting operators, which correspond to discrete translations in phase space along the position and momentum quadratures. These stabilizers impose a lattice structure in phase space, making the code resilient to small displacement errors in both quadratures. In the ideal limit, the logical states of the square-lattice GKP code consist of infinitely many infinitely squeezed states, with each component represented by a Dirac delta function. The two logical states of the square lattice GKP code are given by
\beq\label{eq:idealgkp}
|\bar{0}_L\rangle_{\rm GKP}^{\rm ideal} = \sum_{s \in \mathbb{Z}} |q = 2s\sqrt{\pi} \rangle, \quad |\bar{1}_L\rangle_{\rm GKP}^{\rm ideal} = \sum_{s \in \mathbb{Z}} |q = (2s + 1)\sqrt{\pi} \rangle.
\eeq
However, these states are non-normalizable and therefore unphysical, as they would require infinite energy to realize in practice.

In view of this, one typically considers a finite-energy approximation of the GKP code. This code is characterized by a real width parameter $\sigma \in [0,1]$ and can be expressed in the Fock basis as  
\begin{equation}
\label{eq:finiteGKP}
\ket{\psi_{\mathrm{GKP}}^{\sigma,\mu}} = \sum_{\alpha \in \mathcal{K}(\mu)} 
e^{-\sigma^2|\alpha|^2} e^{-i \mathrm{Re}[\alpha] \mathrm{Im}[\alpha]} 
D(\alpha) \ket{0},
\end{equation}
where $\hat{D}(\alpha)$ is the displacement operator defined as $\hat{D}(\alpha) = \exp(\alpha \hat{a}^\dagger - \alpha^\ast \hat{a})$, and $\alpha$ takes values on a discrete in phase space: 
\begin{equation}
\mathcal{K}(\mu) = \sqrt{\frac{\pi}{2}} \,(2n_1 + \mu) 
+ i\sqrt{\frac{\pi}{2}} \,n_2, \quad n_1, n_2 \in \mathbb{Z}.
\end{equation}
Here $\mu \in \{0,1\}$ labels the logical states, corresponding to a relative displacement of $\sqrt{\pi/2}$ along the position ($q$) quadrature when $\mu=1$ compared to $\mu=0$. The Gaussian envelope $e^{-\sigma^2|\alpha|^2}$ introduces an energy cutoff, regularizing the otherwise ideal, infinite-energy GKP code. The phase factor $e^{-i \mathrm{Re}[\alpha] \mathrm{Im}[\alpha]}$ ensures the correct symplectic structure in phase space, maintaining consistency with the underlying canonical commutation relations.

In this case, the logical states $\ket{\bar{0}_L}_{\rm GKP}^{\rm finite}$ and $\ket{\bar{1}_L}_{\rm GKP}^{\rm finite}$ are defined as  
\beqa\label{eq:approxigkp}
\left \{ \begin{array}{lll}
|\bar{0}_L\rangle_{\rm GKP}^{\rm finite} = \ket{\psi_{\mathrm{GKP}}^{\sigma,\mu=0}},\\
|\bar{1}_L\rangle_{\rm GKP}^{\rm finite} = \ket{\psi_{\mathrm{GKP}}^{\sigma,\mu=1}}.
\end{array}\right.
\eeqa
These two states constitute an approximately orthogonal basis in the logical space, with an overlap $\sim e^{-\pi/(4\sigma^2)}$, which decays exponentially with increasing squeezing strength. As $\sigma \to 0$, the ideal GKP limit is recovered, and the codewords become exactly orthogonal.

Despite their promising features, the experimental realization of GKP codes remains highly challenging. It requires precise control over non-Gaussian operations, such as ancilla-assisted measurements, photon subtraction, and engineered interactions in circuit-QED or trapped-ion architectures. In addition, the implementation demands a high level of squeezing (often exceeding 10 dB), as well as ultra-low-loss and high-coherence environments, which impose stringent requirements on quantum hardware. Although recent experiments have successfully demonstrated approximate finite-energy GKP states in superconducting resonators~\cite{gkp2025cQED} and trapped-ion systems~\cite{gkp2022ions}, realizing fault-tolerant quantum computation based on GKP codes remains an open and demanding frontier.

\begin{figure}[h]
\centering
\includegraphics[scale=0.6]{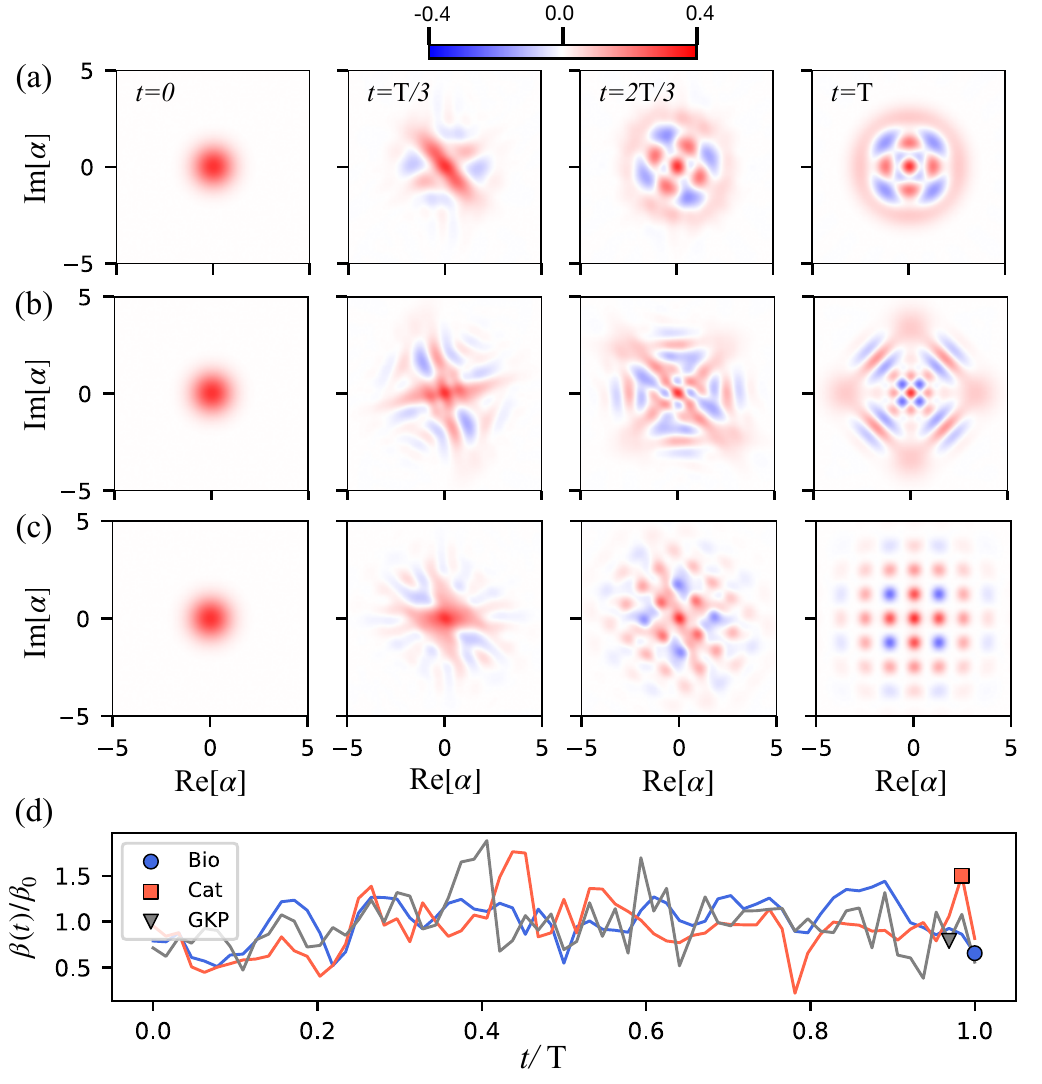} %%% width =\columnwidth scale=0.3
\caption{
\textbf{Bosonic code-state preparation via optimal pulse engineering (OPE).} 
(a-c) Snapshots of the Wigner functions during state preparation from vacuum to target bosonic code states using the optimal control technique for (a) binomial, (b) cat, and (c) GKP codes, respectively. (d) Piecewise trajectories of the optimized control parameter $ \beta(\tau)$, normalized by $\beta_0$, for the three bosonic codes. All final infidelities approach $10^{-5}$.  
Parameters: $N_k=50$, $\sigma=0.35$, $N_t=64$, $d=32$, $\delta =1$, and $\alpha=2.3447$.
}
\label{fig:figure_code}
\end{figure}

\section{Appendix F: Quantum optimal control}
In this section, we provide details of the quantum optimal control technique used in the main text. The control problem is formulated as the constrained minimization of a loss function
\begin{equation}
\vec{\beta}^{\rm opt} = \arg\min_{\vec{\beta}} \mathcal{L}(\vec{\beta})
\quad \text{s.t.} \quad | \vec{\beta}-\beta_0 | \le \delta,\quad \delta\in\mathbb{R}^+,
\end{equation}
where the loss function $\mathcal{L}(\vec{\beta})$ is task specific, and the controller $\vec{\beta}$ parametrizes the piecewise-constant amplitude $\beta(t)$, which is bounded by a predefined power constraint $\delta$. In practice, we use the state (gate) infidelity $1-\mathcal{F}_{\rm state,(gate)}$ as the loss function for target-state preparation (single-logical-gate synthesis).
More specifically, the state fidelity is defined as $\mathcal{F}_{\rm state} = |\langle \psi_{\rm fin} | \psi_{\rm tar} \rangle|^2$ between the final and target states, and is used for Haar-random state preparation and bosonic code-state preparation. For logical gate synthesis, we instead use the gate infidelity $1-\mathcal{F}_{\rm gate}$, where $\mathcal{F}_{\rm gate}$ is defined in Eq.~(12) of the main text, as the optimization objective.
To solve this optimization problem, we employ a gradient-based constrained optimizer, such as {\tt SLSQP}, implemented in the {\tt Python}-based package {\tt SciPy}~\cite{2020SciPy-NMeth}, which allows us to enforce the required control bounds. For a given Trotter depth $N_t$, the control parameters are randomly initialized within the prescribed range.

% $\vec{\beta}_0 \in \mathbb{R}^{N_t}$

The overall performance of the optimized quantum lattice gates (QLGs) depends on two key factors: Trotter error suppression and optimization efficiency. The former is determined by the Trotter depth, while
the latter is influenced by both the number of control parameters and the imposed boundary conditions. Increasing $N_t$ improves accuracy by reducing
the Trotter error and expanding the optimization landscape. However, this may also decrease the optimization efficiency, since the number of optimization parameters scales as $\mathcal{O}(N_t)$. Therefore, in practice, $N_t$ should be carefully chosen to balance the error suppression and optimization efficiency.  
We note that alternative optimizers, such as other advanced quantum control techniques~\cite{2020SciPy-NMeth} and the gradient-free CRAB algorithm~\cite{CRAB2022one}, can also be applied to our framework. In this work, we adopt the ${\tt SciPy}$ implementation for its practical simplicity and reliable performance in our numerical simulations.
% \hty{One thing might be interesting to discuss: the loss function for the optimization of a logical gate could be the infidelity of physical unitary or the target gate infidelity, where we used the latter one.}

In Figs.~\figpanel{fig:figure_code}{a}--\figpanel{fig:figure_code}{c}, we illustrate the Wingner-function snapshots of the quantum state evolution for preparing the three representative bosonic codes---binomial, cat, and GKP codes---using the optimal control technique mentioned above. Each preparation process starts from the vacuum state and evolves toward the desired logical state under a bounded control field $\vec{\beta}$. As shown in Fig.~\figpanel{fig:figure_code}{d}, the optimized control pulses can be obtained within a few minutes using the gradient-based optimization method. With these optimized pulses, the system eventually evolves to the target states with final infidelities below $10^{-5}$. This demonstrates the efficiency of our optimization framework for high-fidelity preparation of various bosonic code states.

\color{black}

\section{Appendix G: Single-qubit logical gate construction with bosonic codes}

As mentioned above, for bosonic codes, a logical qubit is typically represented by two mutually orthogonal codewords
\begin{equation}
    \ket{\bar{0}_L},\;\ket{\bar{1}_L} \;\in\; \mathcal{H}_\text{Fock},
\end{equation}
where $\mathcal{H}_\text{Fock}$ denotes the truncated Fock space of the bosonic mode.  
Given these codewords in the Fock basis, our goal is to construct an arbitrary logical single-qubit unitary
\begin{equation}
    \hat{U}_L \in \mathbb{SU}(2),
\end{equation}
as a logical unitary acting on the underlying bosonic Hilbert space.

We define the code matrix as
\begin{equation}
    \hat{C} = \big[\, \ket{\bar{0}_L},\, \ket{\bar{1}_L}\,\big] \;\in\; \mathbb{C}^{N\times 2},
\end{equation}
where $N$ is the cutoff dimension of the Fock space.  
Since the codewords, such as those of the GKP code, may not be exactly orthogonal under finite-dimensional truncation, we first construct the overlap matrix
\begin{equation}
    \hat{G} = \hat{C}^\dagger \hat{C},
\end{equation}
and then obtain the corresponding orthonormalized basis as
\begin{equation}
    \hat{Q} = \hat{C}\hat{G}^{-1/2}, \qquad \hat{Q}^\dagger \hat{Q} = \hat{I}_2.
\end{equation}
The projector onto the code subspace is then given by $\hat{P} = \hat{Q} \hat{Q}^\dagger$.  

{The desired embedding of the logical operation $\hat{U}_L$ into the truncated Fock space ($\hat{U}_{\text{phys}}$) is then constructed as
\begin{equation}
\hat{U}_{\text{phys}} = \hat{Q} \hat{U}_L \hat{Q}^\dagger + \left(\hat{I}_N - \hat{P}\right),
\end{equation}
which acts as $\hat{U}_L$ on the code subspace and as the identity on its orthogonal complement.  
Equivalently, let $\hat{B} = \left[\hat{Q}, \hat{Q}_\perp \right]$ be a matrix that extends $\hat{Q}$ to a full unitary matrix. Then $\hat{U}_{\text{phys}}$ can be written in a block-diagonal form as
\begin{equation}
    \hat{U}_{\text{phys}} = \hat{B} \; \mathrm{diag}\left(\hat{U}_L, \hat{I}_{N-2}\right) \hat{B}^\dagger.
\end{equation}
}

To benchmark the quality of an implemented operation $\hat{U}(T)$ relative to the target logical gate $\hat{U}_L$, one evaluates its effective action on the logical subspace:
\begin{equation}
    \hat{U}_{\text{eff}} = \hat{Q}^\dagger \hat{U}(T) \hat{Q}.
\end{equation}
The logical single-qubit gate fidelity is then defined as (see also Eq.~\eqref{eq:gate_fidelity} in the main text)
\begin{equation}
\mathcal{F}_\text{gate} = \frac{1}{6} \left| \mathrm{Tr}\!\left(\hat{U}_L^\dagger \hat{U}_{\text{eff}} \right)\right|^2+\frac{1}{3}.
\end{equation}
This definition is a natural extension of the standard process fidelity to the encoded qubit.  
It measures the normalized Hilbert--Schmidt inner product between the target logical operation and the effective channel realized within the code space. By construction, $\mathcal{F}_\text{gate}=1$ \emph{if and only if} $\hat{U}_{\text{eff}}$ coincides with $\hat{U}_L$ up to a global phase. Any deviation from the ideal behavior, due to control imperfections, leakage, or other errors, reduces the gate fidelity accordingly.

\section{Appendix H: Hardware-aware noise}
%\textit{Decoherence and noise.(could also be in SM)}

To investigate the impact of a noisy environment, we model the system dynamics using the Lindblad master equation in the rotating frame:
\beq
\frac{\partial}{\partial t}\hat{\rho}(t)=-\frac{i}{\lambda}\left[\hat{H}(t), \hat{\rho}(t)\right]+\kappa\mathcal{D}[\hat{a}]\hat{\rho}(t), 
\eeq
where $\hat{\rho}(t)$ is the density operator of the system and $\hat{a}$ is the bosonic annihilation operator. The second term on the right-hand side describes the energy relaxation of the cavity due to single-photon loss at rate $\kappa$, with the Lindblad dissipator defined as $\mathcal{D}[\hat{O}]\hat{\rho}=\hat{O}\hat{\rho}\hat{O}^{\dag}-(\hat{O}^{\dag}\hat{O}\hat{\rho}+\hat{\rho}\hat{O}^{\dag}\hat{O})/2$.

\textit{Incoherent error.} As described above, the dominant decoherence mechanism in many bosonic systems, such as superconducting cavities and trapped-ion oscillators, is energy relaxation through single-photon loss.
For a gate of duration $T$, one can estimate the fundamental fidelity limit imposed by this loss channel in the absence of quantum jumps (i.e., in the no-jump trajectory). Let $\bar{ n }(t)
= \langle \hat{a}^\dagger \hat{a}\rangle(t)$ denote the instantaneous mean photon
number during the unitary (lossless) evolution. Then the probability that no
photon is lost throughout the gate duration is approximately given by
\begin{equation}\label{eq:coherent_limit}
  \mathcal{F}_{\rm gate}(T)\;\lesssim\;
  \mathcal{F}_{\mathrm{coh}}^{\max}(T)=
  \exp\!\left[
    -\kappa \int_0^T \bar{ n }(t)\, dt
  \right].
\end{equation}
This expression provides a simple yet informative upper bound on the achievable gate
fidelity, set solely by the incoherent energy relaxation (i.e., the single-photon loss channel).

{In Fig.~\ref{fig:incoherent_noise}, we show the infidelity of bosonic code-state preparation as a function of the (normalized) single-photon loss rate $\kappa/\omega_{0}$, varied over the range $\kappa/\omega_0 \in [10^{-6}, 10^{-1}]$.
For the bare QLG protocol (blue triangles), the infidelity saturates at the level of $\sim 10^{-2}$ in the limit $\kappa \to 0$, indicating that its performance is predominantly limited by coherent errors such as the rotating-wave approximation and Trotterization.
In contrast, owing to the short total evolution time of our protocol compared to the cavity coherence time, the optimized QLG scheme (opt-QLG, red circles) achieves significantly higher fidelities and closely follows the coherent upper bound given by Eq.~(\ref{eq:coherent_limit}), shown as solid black lines.}

\begin{figure}[h]
\centering
\includegraphics[scale=0.85]{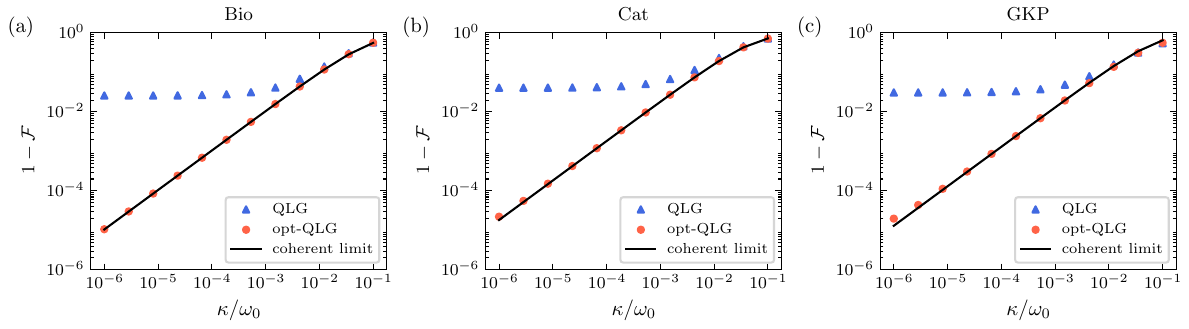} %%% width =\columnwidth scale=0.3
\caption{
\textbf{Incoherent error.} Infidelity of bosonic code-state preparation as a function of the normalized single-photon loss rate $\kappa/\omega_{0}$ for (a) binomial, (b) cat, and (c) GKP codes. Blue triangles denote the infidelities obtained using the standard QLG protocol, while red circles correspond to the optimized QLG (opt-QLG) scheme. The solid black lines show the coherent upper bound given by Eq.~(\ref{eq:coherent_limit}) with the gate duration $T= 2\pi/\omega_0$. All other parameters are the same as those in Fig.~\ref{fig:figure_code}.
}
\label{fig:incoherent_noise}
\end{figure}

\textit{Coherent error.} 
In our setup, coherent errors refer to control inaccuracies induced by imperfections in experimental instrumentation, such as amplitude and frequency deviations. These inaccuracies typically arise in superconducting circuits due to flux noise or distortions in waveform generators.

For a single Josephson junction (JJ) embedded in a SQUID loop with tunable amplitude and phase, the effective driving potential takes the form
\beq\label{eq-squid}
V_{\mathrm{SQUID}}(\hat{\varphi})=-E_J\cos\left(\frac{\Phi_{sq}}{\Phi_0}\pi\right)\cos\left[\frac{M(\Phi_x)}{L}\hat{\varphi}+\frac{\Phi_{ext}}{\Phi_0}\right], 
\eeq
where $\Phi_0$ is the magnetic flux quantum and $M(\Phi_x)/L$ describes the flux-to-phase conversion.
We consider a harmonically modulated external flux (in units of $\Phi_0$) of the form
\beq
\Phi(t) = \Phi_\text{dc} + f(t)\cos\big(\omega_\Phi t + \phi_0\big),
\eeq
where $\Phi_{\rm dc}$ is a static offset, $f(t)$ is a slowly varying modulation envelope, and $\omega_\Phi$ is the carrier frequency.
Note that in realistic experiments, both the DC and AC components are subject to fluctuations. In view of this, we model the noisy external flux as
\beq
\Phi'(t) = \Phi_{\rm dc}+\epsilon_{\rm dc}(t)+ f(t)\cos\big[(\omega_\Phi+\epsilon_{\rm ac}(t)) t + \phi_0\big],
\eeq
where $\epsilon_{\rm dc}(t)$ and $\epsilon_{\rm ac}(t)$ represent the time-dependent amplitude and frequency fluctuations, respectively. {These perturbations result in coherent errors of the effective driving potential, which is expressed as
\beqa 
\hat{V}'(\hat{x},t)=\int_{-\infty}^{+\infty} \beta(t)A'(k, t)\cos[k\hat{x}+\phi'(k,t)]dk. 
\eeqa
}

{
To simulate the effect of coherent noise numerically, we incorporate fluctuations in both the control envelope and the drive frequency by making the replacements 
\beq
\beta(t) \rightarrow \beta(t)+\epsilon_{\rm dc}(t), \quad \omega_\Phi \rightarrow \omega_\Phi+\epsilon_{\rm ac}(t), 
\eeq
where the amplitude noise $\epsilon_{\rm dc}(t)$ and frequency noise $\epsilon_{\rm ac}(t)$ are modeled as Gaussian white-noise processes, i.e.,
 \beq
 \epsilon_{\rm dc/ac}(t) \equiv \zeta_{\rm dc/ac}\xi(t).
\eeq
Here, $\zeta_{\rm dc/ac}$ sets the corresponding noise strength, and $\xi(t)$ denotes standard Gaussian white noise with $\langle\xi(t)\rangle=0$ and $\langle\xi(t)\xi(t')\rangle=\delta(t-t')$.
}

{Figures~\ref{fig:amplitude noise} and \ref{fig:frequency noise} summarize the robustness of our bosonic code-state preparation protocol against coherent control errors for both the standard QLG method and its optimized version (opt-QLG). 
Figure~\ref{fig:amplitude noise} shows the infidelity under amplitude fluctuations with noise strength $\zeta_{\mathrm{dc}}/\beta_0\in[0,0.2]$, while Fig.~\ref{fig:frequency noise} presents the corresponding results under frequency fluctuations with $\zeta_{\mathrm{ac}}/\omega_0\in[0,0.1]$. 
In both cases, the optimized pulses exhibit pronounced robustness, maintaining high fidelity even in the presence of moderate fluctuations, e.g., $\zeta_{\mathrm{dc}}\le 0.1\beta_0$ for amplitude noise and $\zeta_{\mathrm{ac}}/\omega_0 \lesssim 0.02$ for frequency noise. 
These results indicate that the opt-QLG protocol remains resilient under realistic coherent imperfections relevant to superconducting-circuit platforms. 
Moreover, our approach is naturally compatible with robust pulse-engineering techniques that are well developed in superconducting circuits~\cite{robust_pulse}, which can be incorporated straightforwardly to further enhance resilience against coherent errors.}

\begin{figure}[h]
\centering
\includegraphics[scale=0.85]{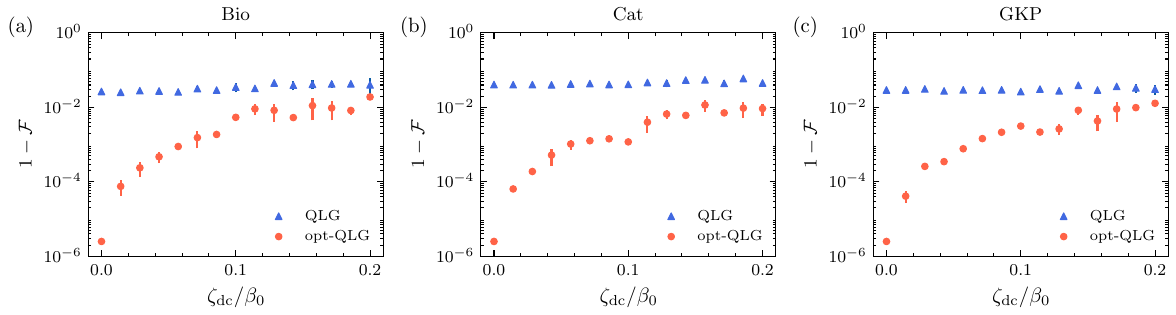} %%% width =\columnwidth scale=0.3
\caption{
{\textbf{Amplitude noise.} Infidelity of bosonic code-state preparation as a function of the normalized amplitude-noise strength $\zeta_{\rm dc}/\beta_0$ for (a) binomial, (b) cat, and (c) GKP codes. The blue triangles denote the results for the standard QLG
protocol, while the red circles correspond to the optimized QLG (opt-QLG) scheme. The error bars indicate the standard deviation over 20 realizations with randomly sampled initial control parameters. All other parameters are the same as in Fig.~\ref{fig:figure_code}.}
}
\label{fig:amplitude noise}
\end{figure}

\begin{figure}[h]
\centering
\includegraphics[scale=0.85]{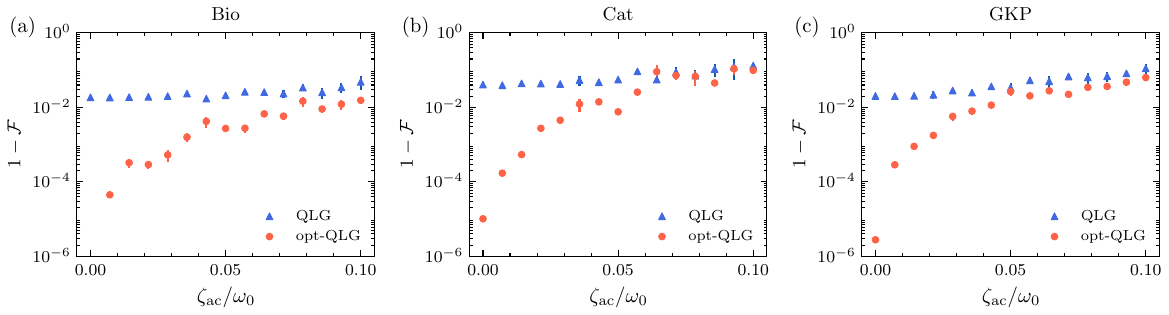} %%% width =\columnwidth scale=0.3
\caption{
{
\textbf{Frequency noise.} Infidelity of bosonic code-state preparation as a function of the normalized frequency-noise strength $\zeta_{\rm ac}/\omega_0$ for (a) binomial, (b) cat, and (c) GKP codes. The blue triangles denote the results for the standard QLG
protocol, while the red circles correspond to the optimized QLG (opt-QLG) scheme. The error bars indicate the standard deviation over 20 realizations with randomly sampled initial control parameters. All other parameters are the same as in Fig.~\ref{fig:figure_code}.}
}
\label{fig:frequency noise}
\end{figure}

\newpage

{
\textbf{\textit{Robustness under composite noise}.}---
To further compare the single-period Floquet (SF) method with conventional adiabatic-ramping (AR) approaches~\cite{LZGuo2025CP,gkpFloqut2024prl}, we benchmark their robustness under realistic noise conditions. 
For the AR protocol, we follow the scheme proposed in Ref.~\cite{LZGuo2025CP} to prepare the binomial logical state $|\bar{0}_L\rangle_{\rm bin}$ [defined in Eq.~\eqref{eq:binomialstates}] starting from the vacuum state, with a total evolution time of $2\times10^3$ Floquet . 
For incoherent errors, the much longer operation time of the AR protocol implies a substantially more stringent coherence requirement than in the SF approach, as also indicated by Eq.~\eqref{eq:coherent_limit}. 
We then consider a more realistic composite-noise model that includes both amplitude and frequency fluctuations.}

{Figure~\ref{fig:figure_robust} shows the infidelity of the prepared binomial code state using the standard SF, optimized SF (opt-SF), and AR approaches as functions of the normalized amplitude- and frequency-noise strengths, $\zeta_{\rm dc}/\beta_0$ and $\zeta_{\rm ac}/\omega_0$, respectively. 
The SF protocols exhibit pronounced robustness, maintaining high fidelity ($>0.99$) for total noise strength below $\sim 10^{-1}$. Here, the total noise strength is simply defined as $\zeta_{\rm tot} = \zeta_{\rm ac}/\omega_0+\zeta_{\rm dc}/\beta_0$.
By contrast, the performance of the AR method deteriorates rapidly when the frequency noise exceeds $\sim 10^{-3}$. 
We attribute this behavior to the combined effect of the frequency ramping and the long evolution time, which makes the AR protocol substantially more sensitive to control noise. 
As mentioned above, robust pulse-engineering techniques~\cite{robust_pulse} can be incorporated to further improve resilience against coherent errors.
}

\begin{figure}[h]
\centering
\includegraphics[scale=0.68]{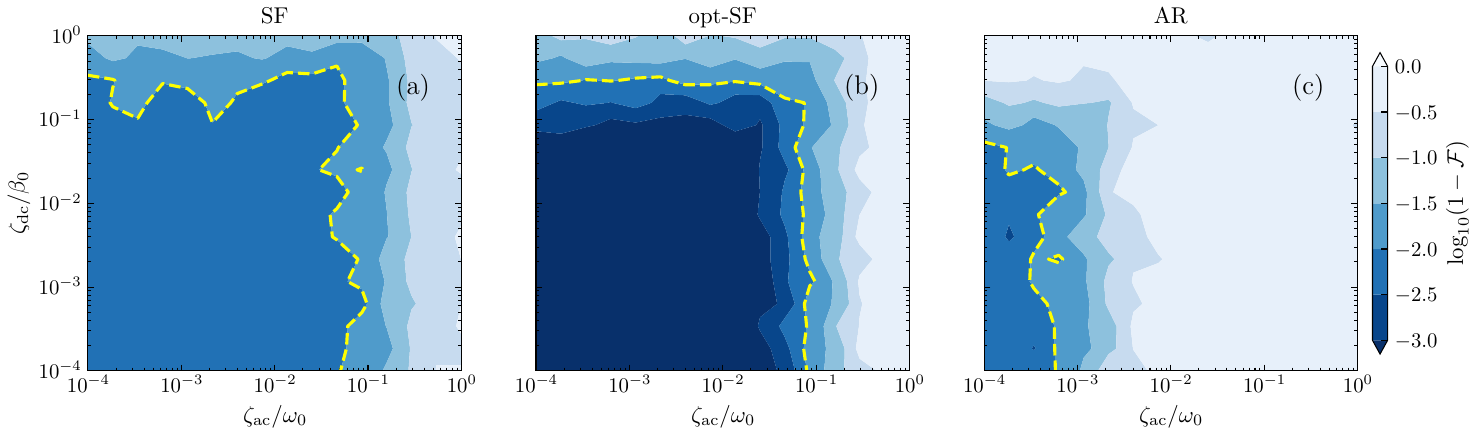} %%% width =\columnwidth scale=0.3
\caption{\label{fig:robustness}
{\textbf{Robustness under composite noise}.
Infidelity of binomial-zero code-state preparation as a function of the normalized frequency noise strength $\zeta_{\rm ac}/\omega_0$ and amplitude noise strength $\zeta_{\rm dc}/\beta_0$ for (a) SF, (b) opt-SF, and (c) AR protocols. 
The yellow dashed curve marks the contour corresponding to an infidelity of $10^{-2}$. 
Each data point is obtained by averaging over 20 numerical realizations. 
For the AR protocol, we use a total evolution time of $2\times10^3$ Floquet periods to prepare the target code state from the vacuum state. 
All other parameters for the SF and opt-SF protocols are the same as those used in Fig.~\ref{fig:figure_code}.
}}
\label{fig:figure_robust}
\end{figure}

% \newpage

{
\section{Appendix I:. Controllability Benchmark}
\textit{\textbf{SNAP-based circuit}}.---To realize universal gate operation on a logical qubit encoded in a quantum oscillator, one typically combines displacement operations with selective number-dependent arbitrary phase (SNAP) gates in a resonator--qubit architecture~\cite{Liang2015praR,heeres2017snap}.  
Since displacement and SNAP operations act on the resonator and the ancillary qubit, respectively~\cite{Liang2015praR}, they are applied sequentially rather than simultaneously.  
This composite operation has become a standard building block for realizing arbitrary unitaries on CV logical qubits, and full $U(d)$ control has been experimentally demonstrated in circuit-QED platforms~\cite{heeres2017snap}.
For simplicity, the composite operation can be written as
\beq
\hat{\mathcal{D}}(\alpha)\hat{S}_n(\vartheta) = e^{i\alpha(\hat{a}^\dagger - \hat{a})} e^{-i \vartheta_n |n\rangle\langle n|},
\eeq
where $\hat{\mathcal{D}}(\alpha)$ is the displacement operator and $\hat{S}_n(\vartheta)$ denotes a SNAP gate with number-dependent phases $\{\vartheta_n\}$. 
Since the universality proof is formulated in the infinitesimal-displacement limit, controllability does not rely on any nonzero lower bound on $|\alpha|$. In practice, however, $|\alpha|$ should be chosen small enough to maintain the validity of a truncated Fock-space description~\cite{Liang2015praR}. 
In the numerics below, we therefore take $\alpha$ in the perturbative regime and use $|\alpha|\sqrt{d}\ll 1$ as a practical guideline to suppress leakage outside the truncated subspace.}

{We consider a SNAP+displacement circuit ensemble acting on the truncated Fock space $\mathcal H_d=\mathrm{span}\{|0\rangle,\dots,|d-1\rangle\}$. Starting from $|0\rangle$, each sample is generated by an $L$-layer circuit
\begin{equation}
\hat{\mathcal{C}}_{\rm snap}=\prod_{\ell=1}^{L} \prod_{j=1}^{d}\hat{D}(\alpha_{\ell}) \hat{S}(\theta_{j,\ell} ),
\end{equation}
where $\alpha\in\mathbb{C}$ is a complex displacement amplitude and $\vartheta_{j,\ell}$ denotes the photon-number-selective phase applied to $|j\rangle$ in layer $\ell$. Each displacement contributes two real trainable parameters (its real and imaginary parts), so the circuit contains 
$L\times (d+2)$ real trainable parameters in total. To generate ensembles with an arbitrary number of trainable parameters, we include all parameters from the first several complete SNAP layers, while any remaining parameters are randomly sampled from the subsequent layer. 
The displacement amplitudes are sampled uniformly from the disk $|\alpha_{\ell}|\le 1/\sqrt{d}$, ensuring operation in the perturbative regime and suppressing leakage outside $\mathcal H_d$.}

{\textit{\textbf{QLG-based circuit}}.---For the QLG ansatz, the gate parameters are obtained analytically. We therefore construct a parametrized circuit by discretizing the control envelope into piecewise-constant values $\beta(t_n)$, yielding 
\begin{equation}
\hat{\mathcal{C}}_{\rm qlg} = \prod_{n=1}^{N_t}{\hat{\mathbf{G}}}[\vec{\beta}(t_n)].
\label{eq:Trotter_error}
\end{equation}
where $N_t$ denotes the number of time-discretization steps (i.e., the Trotter depth). For a given target state, all other QLG parameters are determined uniquely by the analytic construction. Consequently, the only trainable degrees of freedom of the QLG circuit are the discrete control values, which we collect into the vector $\vec{\beta}(t_n)=[\beta_1,\beta_2,...,\beta_{N_t}]$.
}

{
\subsection{Appendix I.1: Finite resource complexity}
To compare different circuit architectures on an equal footing, we quantify the finite computational resource of a circuit ansatz $\mathcal{C}$ by the total number of trainable parameters,
\begin{equation}
\mathcal{R}:=\mathcal{N}[\mathcal{C}(\boldsymbol{\vartheta})],
\end{equation}
where $\mathcal{N}[\cdot]$ denotes the number of free variational parameters in the circuit. 
This definition provides a simple, architecture-independent measure of the available control resources.}

{For the QLG- and SNAP-based circuits, the corresponding resource costs are determined by the numbers of Trotter steps and circuit layers, respectively. 
In particular, $\mathcal{R}_{\rm qlg}$ is set by the total number of free parameters across the $N_t$ QLG steps, while $\mathcal{R}_{\rm snap}$ grows with both the layer number $L$ and the truncated Hilbert-space dimension $d$. 
Therefore, although the QLG ansatz does not depend explicitly on $d$ at the level of circuit construction, its required resource is still controlled by the Trotter depth $N_t$, which must be increased to suppress the Trotter errors.
Since the two protocols differ in their hardware realizations, we adopt the computational resource $\mathcal{R}$, i.e., the number of trainable parameters, as a common metric for benchmarking controllability.
With this definition, we next compare the minimal resource cost required by different circuit architectures to achieve the same state-preparation performance, as quantified by the target fidelity.}

{The resource measure introduced above admits two complementary interpretations. First, it quantifies the \emph{experimental complexity}, since it directly corresponds to the number of tunable operations that must be implemented in the circuit. Second, it characterizes the \emph{numerical controllability} of the circuit ansatz, because the number of free variables strongly affects the optimization performance in a large Hilbert space. From this perspective, the efficiency of the target-state preparation can be naturally assessed by the amount of resource required to reach a prescribed high-fidelity threshold.}

{To further quantify the efficiency of a circuit ansatz, it is useful to introduce a \emph{fixed-error resource measure}. Rather than asking what fidelity can be achieved for a given resource budget, we instead ask how many control parameters are required to reach a prescribed target fidelity. This provides a more operational characterization of controllability, since it directly quantifies the resource cost of high-quality state preparation as the Hilbert-space dimension increases.}

{Specifically, for a target state $|\psi_{\rm tar}\rangle$ and a circuit family $\mathcal{C}$ with resource $\mathcal{R}=N(\mathcal{C})$, we define the largest achievable fidelity at fixed resource as
\begin{equation}
\mathcal{F}^\star_{\mathcal{C}}(\mathcal{R},d)
:=
\max_{\vartheta\in\mathbb{R}^{\mathcal{R}}}
\big|
\langle \psi_{\rm tar} | \hat{U}_{\mathcal{C}}(\vartheta) | \psi_0 \rangle
\big|^2,
\end{equation}
where $\hat{U}_{\mathcal{C}}(\vartheta)$ denotes the unitary implemented by the parametrized circuit acting on the initial state $|\psi_0\rangle$. For a prescribed fidelity threshold $\mathcal{F}_{\rm th}=1-\varepsilon$, we then define the minimal resource cost required to achieve this accuracy as
\begin{equation}
\mathcal{R}_{\min}(d;\varepsilon)
:=
\min\left\{
\mathcal{R}\; \middle|\;
\mathcal{F}^\star_{\mathcal{C}}(\mathcal{R},d)\ge 1-\varepsilon
\right\}.
\end{equation}
Equivalently, one may interpret $\varepsilon$ as the preparation error and study the scaling of $\mathcal{R}_{\min}$ with the Hilbert-space dimension $d$ at fixed error tolerance.}

{This quantity serves as a direct benchmark of circuit efficiency. A slower growth of $\mathcal{R}_{\min}(d;\varepsilon)$ indicates that the ansatz can maintain a prescribed preparation accuracy with fewer resources as the system size increases, whereas a faster growth implies reduced practical controllability. In this way, the dependence of $\mathcal{R}_{\min}$ on $d$ provides a direct and physically meaningful characterization of the resource scaling required for high-fidelity state preparation.
}

{
\subsection{Appendix I.2: Circuit optimization}
In this section, we construct a hybrid quantum-classical algorithm by combining a classical optimizer with a parametrized quantum circuit.
The circuit-optimization problem can be expressed as
\beq
\vartheta_{\rm opt} = \arg\min_{\vartheta} \mathbb{L}[\mathcal{C}(\vartheta)] ,
\eeq
where the loss function $\mathbb{L}[\mathcal{C}(\vartheta)]$ is defined as the state infidelity between the prepared (measured) state and the target state in our setup.
To benchmark the influence of different optimizers, including gradient-free and gradient-based methods, we perform the same optimization task for the QLG and SNAP circuits, respectively.}

{\textit{\textbf{Numerical optimization of SNAP-based state preparation.}}---
For a target state $|\psi_{\rm tar}\rangle$ in a truncated Hilbert space of dimension $d$, we numerically optimize a SNAP-based state-preparation circuit starting from the vacuum state $|0\rangle$. The variational state is written as
\begin{equation}
|\psi(\vec{\vartheta})\rangle
=
\hat{D}(\alpha_L)\hat{S}(\vartheta^{(L)})\cdots
\hat{D}(\alpha_1)\hat{S}(\vartheta^{(1)})|0\rangle,
\end{equation}
where $\hat{S}(\vartheta)=\sum_{n=0}^{d-1}e^{i\vartheta_n}|n\rangle\langle n|$ is the SNAP gate and $\alpha_\ell=x_\ell+i y_\ell$ is the complex displacement amplitude in layer $\ell$. We fix a total number of free real parameters $\mathcal{R}_{\rm total}$ and distribute them sequentially across layers. Concretely, we include all parameters from the first several complete SNAP layers, and, if additional parameters are needed, we sample the remaining parameters from the subsequent layer. Each layer contains at most $d+2$ trainable parameters: $d$ SNAP phases and two displacement quadratures. When a layer is only partially filled, we treat only the lowest-index SNAP phases as active parameters.
The loss function is the state infidelity,
\begin{equation}
\mathbb{L}=1-|\langle\psi_{\rm tar}|\psi(\vec{\vartheta})\rangle|^2,
\end{equation}
which is minimized by using the L-BFGS-B optimizer with box constraints $\theta_n\in[0,2\pi)$ and $x_\ell,y_\ell\in[-\alpha_{\max},\alpha_{\max}]$. We choose $|\alpha_{\max}|=  \sqrt{d}$ to remain in the perturbative regime while allowing a sufficiently broad optimization landscape. The variational parameters are initialized randomly within these bounds. When needed, we employ a multi-start strategy and select the run with the largest final overlap with $|\psi_{\rm tar}\rangle$, following the procedure described in the main-text section ``\textit{Optimal pulse engineering}''.}

{For the QLG circuit, we initialize the controller by randomly sampling the control vector $\vec{\beta}(t_n)$ from the constrained $N_t$-dimensional sphere  $\|\vec{\beta}\|/\beta_0 \in \delta$, while all other parameters are determined analytically by the target state (see the main-text section ``\textit{Single-period Floquet control with QLGs}'').
All numerical optimizations are performed using the {\tt Python} package {\tt SciPy}~\cite{2020SciPy-NMeth}. Unless otherwise stated, we use the \emph{gradient-based optimizer} ${\tt SLSQP}$ and set the convergent criterion to $\sim 10^{-5}$.}

%\bibliographystyle{apsrev4-1}
%\bibliography{reference}	
\end{document}